\documentclass[fleqn,usenatbib]{mnras}

\usepackage[T1]{fontenc}
\usepackage[utf8]{inputenc}
\usepackage{xcolor}

\DeclareRobustCommand{\VAN}[3]{#2}
\let\VANthebibliography\thebibliography
\def\thebibliography{\DeclareRobustCommand{\VAN}[3]{##3}\VANthebibliography}

\usepackage{acro}
\usepackage{units}
\usepackage{booktabs}
\usepackage{dcolumn}
\usepackage{makecell}
\usepackage{graphicx}
\usepackage{tabularx}
\usepackage{amsmath,amsfonts,amssymb}
\usepackage{xspace}
\usepackage{mathrsfs}
\usepackage[normalem]{ulem}
\usepackage{multirow}
\usepackage{color}
\usepackage{natbib}
\usepackage{verbatim}
\usepackage{physics}
\usepackage{braket}
\usepackage{colortbl}
\usepackage{newtxtext,newtxmath}

\DeclareOldFontCommand{\rm}{\normalfont\rmfamily}{\mathrm}
\DeclareOldFontCommand{\sf}{\normalfont\sffamily}{\mathsf}
\DeclareOldFontCommand{\tt}{\normalfont\ttfamily}{\mathtt}
\DeclareOldFontCommand{\bf}{\normalfont\bfseries}{\mathbf}
\DeclareOldFontCommand{\it}{\normalfont\itshape}{\mathit}
\DeclareOldFontCommand{\sl}{\normalfont\slshape}{\@nomath\sl}
\DeclareOldFontCommand{\sc}{\normalfont\scshape}{\@nomath\sc}
\DeclareRobustCommand*\cal{\@fontswitch\relax\mathcal}
\DeclareRobustCommand*\mit{\@fontswitch\relax\mathnormal}

\AtBeginDocument{\mathcode`v=\varv}

\DeclareAcronym{EOS}{short=EOS,long=equation of state}
\DeclareAcronym{GW}{short=GW,long=gravitational wave}
\DeclareAcronym{BH}{short=BH,long=black hole}
\DeclareAcronym{NS}{short=NS,long=neutron star}
\DeclareAcronym{BHNS}{short=BHNS,long=black hole - neutron star}
\DeclareAcronym{BNS}{short=BNS,long=binary neutron star}
\DeclareAcronym{NSE}{short=NSE,long=nuclear statistical equilibrium}
\DeclareAcronym{LTE}{short=LTE,long=local thermodynamic equilibrium}
\DeclareAcronym{KN}{short=KN,long=Kilonova}
\DeclareAcronym{EM}{short=EM,long=electromagnetic}

\definecolor{cyan}{rgb}{0,0.9,0.9}
\definecolor{orange}{rgb}{0.9,0.5,0}
\definecolor{magenta}{rgb}{1,0,1}
\definecolor{purple}{rgb}{0.8,0.4,0.8}
\definecolor{darkgreen}{rgb}{0.0,0.5,0.0}
\definecolor{gray}{rgb}{0.8242,0.8242,0.8242}
\definecolor{cadmiumgreen}{rgb}{0.0, 0.42, 0.24}
\definecolor{olive}{rgb}{0.5, 0.5, 0.0}

\newcommand{\reftab}[1]{{Table~\ref{#1}}}
\newcommand{\refsec}[1]{{Sec.~\ref{#1}}}
\newcommand{\reffig}[1]{{Fig.~\ref{#1}}}
\newcommand{\refapp}[1]{{Appendix~\ref{#1}}}
\newcommand{\refeq}[1]{{Eq.~(\ref{#1})}}

\newcommand{\LSm}{LS220$^\dagger$\@\xspace}
\newcommand{\LSl}{LS175$^\dagger$\@\xspace}
\newcommand{\LSh}{LS255$^\dagger$\@\xspace}
\newcommand{\msm}{$m^*_{0.8}$\@\xspace}
\newcommand{\msS}{$m^*_{\textnormal{S}}$\@\xspace}
\newcommand{\msKS}{$(m^* K)_{\textnormal{S}}$\@\xspace}

\title[Nuclear matter properties in BNS mergers]{Impact of nuclear matter properties on the nucleosynthesis and the kilonova from binary neutron star merger ejecta}

\author[G. Ricigliano et al.]{Giacomo Ricigliano$^{1}$\thanks{Contact e-mail: giacomo.ricigliano@gmail.com},
Maximilian Jacobi$^{2}$
and Almudena Arcones$^{1,3,4}$
\\
$^{1}$Institut für Kernphysik, Technische Universität Darmstadt, Schlossgartenstr. 2, Darmstadt 64289, Germany\\
$^{2}$Theoretisch-Physikalisches Institut, Friedrich-Schiller Universit{\"a}t Jena, Jena 07743, Germany\\
$^{3}$GSI Helmholtzzentrum für Schwerionenforschung GmbH, Planckstr. 1, Darmstadt 64291, Germany\\
$^{4}$Max-Planck-Institut für Kernphysik, Saupfercheckweg 1, Heidelberg 69117, Germany}

\date{Accepted XXX. Received YYY; in original form ZZZ}

\pubyear{2024}

\begin{document}
\label{firstpage}
\pagerange{\pageref{firstpage}--\pageref{lastpage}}
\maketitle

\begin{abstract}
Material expelled from binary neutron star (BNS) mergers can harbor r-process nucleosynthesis and power a Kilonova (KN), both intimately related to the astrophysical conditions of the ejection.
In turn such conditions indirectly depend on the equation of state (EOS) describing matter inside the neutron star.
Therefore, in principle the above observables can hold valuable information on nuclear matter, as the merger gravitational wave signal already does.
In this work, we consider the outcome of a set of BNS merger simulations employing different finite-temperature nuclear EOSs.
The latter are obtained from a Skyrme-type interaction model where nuclear properties, such as the incompressibility and the nucleon effective mass at saturation density, are systematically varied.
We post-process the ejecta using a reaction network coupled with a semi-analytic KN model, to asses the sensitivity on the input EOS of the final yields and the KN light curves.
Both of them are found to be non-trivially influenced by the EOS, with the overall outcome being dominated by the heterogeneous outflows from the remnant disk, hosting a variable degree of neutron-rich material.
The dynamical ejecta can be more directly related to the EOS parameters considered, however, we find its role in the yields production and the KN emission too entangled with the other ejecta components, in order to infer solid correlations.
This result highlights the strong degeneracy that intervenes between the merger outcome and the behaviour of the intrinsic nuclear matter, and places itself as a limit to the employment of EOS-constraining approaches of such kind.
\end{abstract}


%%%%%%%%%%%%%%%%%%%%%%%%%%%%%%%%%%%%%%%%%%%%%%%%%%%%%%%%%%%%%%%%%%%%%%%%%%%%%

%% ==============================================================
%%
%%                        INTRODUCTION
%%
%% ==============================================================

\section{Introduction}

The recently born multi-messenger era, initiated by the first joint \ac{GW} and \ac{EM} detection of a \ac{BNS} merger \citep{TheLIGOScientific:2017qsa,GBM:2017lvd,Abbott:2018wiz}, has opened an independent astronomical window on the study of nuclear matter, in addition to direct \ac{NS} measurements such as the mass-radius observations from NICER \citep{Miller:2019cac,Riley:2019yda,Miller:2021qha,Riley:2021,Ludlam:2022,Salmi:2022cgy}.
Many of the observable features of \ac{BNS} mergers are strongly connected to the \ac{EOS} of matter at the high densities found inside the \ac{NS}, where typical values exceed nuclear saturation density.
The merger \ac{GW} signal encapsulates information on the masses of the original \acp{NS} and their tidal deformability, which are intimately linked to the \ac{EOS} \citep[see, e.g.][]{Flanagan:2007ix,Hinderer:2009ca,Hotokezaka:2016bzh,TheLIGOScientific:2017qsa,Annala:2017llu,Tews:2018chv,Abbott:2018exr}.
Both the \ac{GW} and \ac{EM} emissions provide insight on the evolution of the merger remnant, which is strongly affected by the internal structure of the \ac{NS} \citep[see, e.g.,][]{Bauswein:2012ya,Ravi:2014,Bernuzzi:2015rla,Radice:2016rys,Margalit:2017dij,Shibata:2017xdx,Bauswein:2017vtn,Rezzolla:2017aly,Radice:2017lry}.
Furthermore, among the \ac{EM} transients, the \ac{KN} reflects the properties of the radioactive material expelled in the process \citep{Li:1998bw,Metzger:2010sy}, which depend on the thermodynamic evolution of the nuclear matter powering the emission.
These dependencies also propagate to the study of the chemical evolution of the galaxy, to which the final yields of merger events are a relevant input \citep[see, e.g.,][]{Thielemann:2017acv,Cote:2018qku}.

The quantitative study of \ac{BNS} mergers requires general-relativistic 3-dimensional hydrodynamical simulations, where a model for the high density \ac{EOS} is necessary to describe the fluid pressure.
The latter can have a major role in determining for example if and when the central remnant will collapse into a \ac{BH}, how massive and extended the accretion torus will be, and the amount and composition of material expelled during different phases of the system's evolution \citep{Sekiguchi:2016bjd,Bovard:2017mvn,Radice:2018pdn,Radice:2018xqa,Nedora:2020pak,Nedora:2020qtd,Camilletti:2022jms}.
These outcome details are then of fundamental importance for the nucleosynthesis harbored in the merger ejecta as well as for the subsequent \ac{KN} emission.
On one hand, the nuclear burning depends on the thermodynamics of the outflow, whereas for example a cold environment rich enough in neutrons such as the tidal ejecta sets the stage for a strong r-process to occur \citep[][for a recent review on r-process]{Lattimer:1974a,Eichler:1989ve,Freiburghaus:1999a,Rosswog:2017sdn,Cowan:2019pkx}.
The yields of such event, which are typically calculated using a nuclear reaction network involving thousands of nuclear species \citep[see, e.g.,][]{Winteler:2012hu,Lippuner:2017tyn,Reichert:2023}, are thus affected by which outflow component dominates the merger ejection.
On the other hand, the \ac{KN} powered by the radioactive decay of the r-process elements, broadly depends on the overall mass of the ejecta, their opacity to optical photons, which is a function of the composition, their expansion velocity, and their spatial distribution \citep[see, e.g.,][for a recent review]{Arnett:1982,Pinto:2000,Barnes:2013wka,Tanaka:2013ana,Metzger:2014ila,Perego:2014qda,Martin:2015hxa,Perego:2017wtu,Rosswog:2017sdn,Wollaeger:2017ahm,Kasen:2018drm,Metzger:2019zeh}.
The physics behind these dependencies is captured to different degrees of accuracy by a variety of \ac{KN} models.
Already only the first days of \ac{KN} emission (when the ejecta is optically thick and it can be assumed to be in a \ac{LTE} condition) can be synthetically reproduced through detailed radiative transfer simulations \citep[see, e.g.,][]{Tanaka:2013ana,Kasen:2017,Wollaeger:2017ahm,Bulla:2019muo,Bulla:2023,Shingles:2023kua}, radiation hydrodynamics simulations \citep{Morozova:2015bla,Wu:2021ibi,Magistrelli:2024}, or more approximated speed-oriented models \citep[see, e.g.,][]{Grossman:2013lqa,Martin:2015hxa,Hotokezaka:2020,Ricigliano:2024}.

Several works have explored the impact of the nuclear \ac{EOS} model in \ac{BNS} merger simulations, in some cases extending the analysis also to the composition of the ejected material and to the \ac{KN} emission.
The finite-temperature, composition-dependent \acp{EOS} considered were typically obtained from a pool of different nuclear models, derived from relativistic mean field and Skyrme density functional models \citep[see, e.g.,][]{Lattimer:1991nc,Steiner:2012rk,daSilvaSchneider:2017jpg,Bombaci:2018ksa}.
Theses studies, therefore, varied many different \ac{EOS} properties at the same time, and thus mostly assessed the impact of the degree of stiffness of the \ac{EOS}, typically found to be degenerate with other characterizing merger parameters, such as the original binary mass ratio \citep[see, e.g.,][]{Neilsen:2014hha,Sekiguchi:2016bjd,Bovard:2017mvn,Radice:2017lry,Coughlin:2018miv,Radice:2018pdn,Most:2018eaw,Perego:2019adq,Dietrich:2020efo,Hammond:2021vtv,Nedora:2021eojb,Perego:2021mkd,Camilletti:2022jms}.
More systematic investigations of specific \ac{EOS} properties were carried out with different levels of sophistication in, e.g., \citet{Bauswein:2010dn,Raithel:2019gws,Fields:2023,Blacker:2024}, which focused on thermal effects, or \citet{Most:2021ktk}, focusing on the nuclear symmetry energy.

This work is intended as a reprise of the study conducted by \citet{Jacobi:2023}, where selected nuclear matter properties characterizing the nuclear \ac{EOS} were varied systematically, and their impact on the merger dynamics, mass ejection, and \ac{GW} emission was investigated.
The study focused on the variation of the incompressibility and of the effective nucleon mass, which can affect the merger evolution on different levels, since both the cold and the thermal part of the \ac{EOS} are modified.
Here, we extend the analysis further, focusing on the nucleosynthesis harbored in the merger ejecta, and predicting synthetic light curves of the subsequent \ac{KN} transient.
The paper is organized as follows.
We describe the initial setup and the framework used to post-process data from the merger simulations in \refsec{sec:methods}.
In \refsec{sec:eos_models}, we list the \ac{EOS} models considered.
We then describe the procedure for the extraction of fluid elements in \refsec{sec:ejecta_extraction}, the setup for the nucleosynthesis calculations in \refsec{sec:nucleosynthesis_calculations}, and the \ac{KN} semi-analytic model in \refsec{sec:kilonova_model}.
We dedicate \refsec{sec:results} to the analysis of the general properties of the ejecta (\refsec{sec:ejecta_properties}), their final composition (\refsec{sec:nucleosynthesis_yields}), and the corresponding \ac{KN} light curves (\refsec{sec:kilonova_lightcurves}).
The discussion on the reliability of our results is then extended in \refsec{sec:extension}.
Finally, the main findings are reported, together with some final remarks, in \refsec{sec:conclusion}.

%% ==============================================================
%%
%%                           METHODS
%%
%% ==============================================================

\section{Methods}
\label{sec:methods}

This work is based on a series of 3D general-relativistic hydrodynamical simulations published in \citet{Jacobi:2023}.
All the simulations follow the merger of a symmetric binary neutron star system with masses $1.365~M_\odot$.
A different nuclear \ac{EOS} was used for each simulation (see \refsec{sec:eos_models}).
The simulations were performed using the \texttt{WhiskyTHC} code \citep{Radice:2012cu} within the \texttt{EinsteinToolkit} suite \citep{Haas:2020a}.
Neutrino emission and absorption were modeled by a grey leakage scheme and a one-moment neutrino transport scheme, respectively \citep{Radice:2018pdn}.
For further details on the simulations setup, we refer the reader to \citet{Jacobi:2023}.

\subsection{\ac{EOS} models}
\label{sec:eos_models}

\begin{table*}
    \centering
    \caption{List of employed \ac{EOS} models and their nuclear matter properties at saturation density.
    The value of the effective nucleon mass is given in terms of the nucleon mass $m_{\rm n}=939.5654$ MeV.
    We also report the corresponding values of the reduced tidal deformability $\tilde{\Lambda}$ and \ac{NS} radii of initial binaries.
    The model \LSl is highlighted in grey since it is excluded from our analysis due to the prompt collapse of the merger remnant.
    Red and blue values indicate the models used to analyse the systematic variation of the effective nucleon mass and the incompressibility, respectively.}
    \begin{tabular}{l||cccccc|cc}
    \toprule
    \ac{EOS} model & $m^*$         & $B$   & $K$   & $E_{\rm sym}$ & $L$   & $\rho_0$                    & $\tilde{\Lambda}$ & $R_{\textnormal{NS}}$ \\
              & [$m_{\rm n}$] & [MeV] & [MeV] & [MeV]         & [MeV] & [$10^{14}~{\rm g~cm^{-3}}$] &                   & [km]                   \\
    \midrule
    \rowcolor{gray!50} \LSl   &                    1.0   & 16.0 &                     175 & 29.3 &  73.7 & 2.57 &  358.9 & 12.1 \\
                       \LSm   & \cellcolor{red!50} 1.0   & 16.0 & \cellcolor{blue!50} 220 & 29.3 &  73.7 & 2.57 &  606.2 & 12.7 \\
                       \LSh   &                    1.0   & 16.0 & \cellcolor{blue!50} 255 & 29.3 &  73.7 & 2.57 &  661.1 & 13.0 \\
                       \msm   & \cellcolor{red!50} 0.8   & 16.0 &                     220 & 29.3 &  79.3 & 2.57 &  698.4 & 12.9 \\
                       \msS   & \cellcolor{red!50} 0.634 & 16.0 & \cellcolor{blue!50} 220 & 29.3 &  86.5 & 2.57 &  765.4 & 13.2 \\
                       \msKS  &                    0.634 & 16.0 & \cellcolor{blue!50} 281 & 29.3 &  86.5 & 2.57 &  975.0 & 13.5 \\
                       SkShen &                    0.634 & 16.3 &                     281 & 36.9 & 109.4 & 2.41 & 1295.5 & 14.5 \\
                       Shen   &                    0.634 & 16.3 &                     281 & 36.9 & 110.8 & 2.41 & 1220.8 & 14.5 \\
    \bottomrule
    \end{tabular}
    \label{tab:eos}
\end{table*}

We consider seven \acp{EOS} calculated using the \texttt{SROEOS} code \citep{daSilvaSchneider:2017jpg}.
The latter provides \acp{EOS} based on a Skyrme functional, tabulated as a function of density, temperature, and electron-fraction.
Of the seven \acp{EOS}, five were first used in \citet{Yasin:2020a} and two were newly created in \citet{Jacobi:2023}.
In the \acp{EOS} computation, the nuclear matter properties at saturation density were sistematically varied, specifically the incompressibility $K$, the effective nucleon mass $m^*$, the symmetry energy $E_{\rm sym}$, the binding energy $B$, and the saturation density $n_0$.
Furthermore, the Shen \ac{EOS} \citep{Shen:1998gq} is considered for the sake of comparison.
In \reftab{tab:eos}, we summarize the \ac{EOS} models used in this work. We report the values of the nuclear matter properties, together with the tidal deformability and radius of the cold non-rotating neutron stars from the initial binary.
We exclude the model \LSl from our analysis, since this leads to prompt collapse of the merger remnant and thus negligible amount of ejecta.

The fiducial model \LSm is based on the L\&S \ac{EOS} with an incompressibility value of \unit[220]{MeV} \citep{Lattimer:1991nc}.
First, the incompressibility is varied within uncertainties predicted by chiral effective field theory calculations \citep{Hebeler:2011, Drischler:2016, Drischler:2019}, leading to the models \LSl and \LSh.
Subsequently, the effective mass is changed from the \LSm value of $m^*=m_{\rm n}$, where $m_{\rm n}$ is the nucleon mass, to $m^*=0.8~m_{\rm n}$ (\msm) and $m^*=0.634~m_{\rm n}$ (\msS), which is the value found in the Shen \ac{EOS}.
The remaining nuclear matter properties are then modified to match the corresponding values of the Shen \ac{EOS}, starting from the model \msS.
The matched parameters include first $K$, and then $E_{\rm sym}$, $n_0$ and $B$, leading to the models \msKS and SkShen respectively \citep[see][for a more detailed overview of the \ac{EOS} models]{Yasin:2020a,Jacobi:2023}.

As found by \citet{Jacobi:2023}, the incompressibility affects the slope of the cold pressure around saturation density, with higher $K$ values leading to steeper pressure-density profiles.
The effective nucleon mass instead impacts both the cold and the thermal contributions to the pressure.
Within this framework, a lower effective mass leads to a larger slope parameter $L$ and therefore an increase of the \ac{EOS}' stiffness at lower densities.
Furthermore, lower $m^*$ values, increase the thermal contribution to the pressure.

\subsection{Ejecta extraction}
\label{sec:ejecta_extraction}

In order to perform the nucleosynthesis calculations, we follow the ejected matter using Lagrangian tracer particles.
We obtain their trajectories by integrating the fluid velocity fields backwards in time until they reach a temperature above \unit[10]{GK}.
The integration is performed in a post-processing step based on the simulation output.
For an overview of the advantages and disadvantages of backward tracer particle evolution see, e.g., \citet{Sieverding:2023}, while for an assessment on the impact of the neglected dynamics-nucleosynthesis coupling, see \citet{Magistrelli:2024}.

The tracer particles are separated into two sets.
\textit{Set 1} is initialized on a spherical surface with a radius of $R \approx \unit[1500]{km}$, in regular time intervals of $\Delta t \approx \unit[1]{ms}$, in order to record matter that exits the domain boundaries within the simulated time.
A similar scheme was used in \citet{Fujibayashi:2020dvr} but for 2D simulations.
\textit{Set 2} is initialized inside the spherical volume at the end of the simulation, to account for ejected matter that did not reach the outer boundary by the end of the simulation.
Tracer particles are only created where the matter fulfills the Bernoulli criterion, i.e. in the fluid element satisfying the condition $-hu_t>h_\infty$, where $h=1+\epsilon+P/\rho$ is the relativistic specific enthalpy, $u_t$ is the time component of its four-velocity and $h_\infty={\rm lim}_{\rho, T\rightarrow0}h$ is the asymptotic specific enthalpy \citep[see, e.g.,][for an overview of ejection criteria]{Foucart:2021}.

The initial distribution of tracer particles is critical to ensure that the fluid motion is sufficiently well resolved at all times relevant for the nucleosynthesis \citep[see, e.g.,][for an overview of different tracer particle distribution schemes]{Bovard:2017}.
In the following, we outline the method that we use to set the starting points for the backward evolution of our tracer particles.

For \textit{set 1}, we divide the spherical surface into a rectangular grid where each cell has the same solid angle.
Similarly, we divide the sphere of \textit{set 2} into cubic cells of equal volume.
Next, we calculate the mass contained in each cell.
For \textit{set 1}, the masses are calculated by integrating the radial rest-mass flux over the cell area:
\begin{equation}\label{eq:mass_set_1}
    m_1 = R^2 \Delta t \int_{\text{cell}} \rho V^r \, \mathrm{d}\Omega \, ,
\end{equation}
where the radial fluid velocity is given by $V^r = \alpha v^r - \beta^r$, with $\alpha$ the lapse function, $v^r$ the radial 3-velocity and $\beta^r$ the radial shift-vector.
We split each cell with a mass greater than a maximum mass $m_{\rm max} \approx 10^{-6}~M_\odot$ into 4 smaller cells with equal solid angles.
This procedure is repeated for all timesteps until all cells contain less mass than $m_{\rm max}$.
The masses for \textit{set 2} are instead given by the volume integral of the rest-mass density over the cell volume:
\begin{equation}\label{eq:mass_set_2}
    m_2 = \int_{\text{cell}} \rho \, \mathrm{d}V \, .
\end{equation}
Equivalently to \textit{set 1}, we split all cells with $m_2 > m_{\rm max}$ into 8 smaller cells with equal volume until all cells have a mass below $m_{\rm max}$.
Finally, a tracer particle is placed at a random location in each cell in order to cancel the bias introduced by the original regular grid.
This procedure ensures that the mass of all tracer particles is small ($< 10^{-6}~M_\odot$), while at the same time the regions of very low rest-mass density are not underresolved, since the minimum spatial resolution is given by the initial grid.

In \citet{Jacobi:2023}, the ejected matter is classified on the basis of its properties at an extraction radius of \unit[300]{km}.
The dynamical ejecta are defined as the material fulfilling the geodesic criterion $-u_t>1$, as in \citet{Nedora:2020pak,Combi:2023}.
Such ejecta are further divided depending on their electron fraction, whereas fluid elements with $Y_e \leq 0.1$ are classified as tidal ejecta and fluid elements with $Y_e > 0.1$ as shock-heated ejecta.
Finally, matter that fulfills the Bernoulli but not the geodesic criterion is categorized as disk ejecta.
We find that the resulting distinction generally agrees with those obtained by employing different criterions, such as separating the ejecta according to a defined ejection time or on the basis of the remnant density evolution.
Therefore, here we adopt the same classification scheme in order to be consistent across the two studies.

Note that in the previous analysis from \citet{Jacobi:2023}, the extraction radius is used to identify the matter ejected, while here the ejection criterion is evaluated at the latest possible time for each fluid element.
Therefore, the total ejecta mass reported in this work is slightly different and should be more exact.

\subsection{Nucleosynthesis calculations and heating treatment}
\label{sec:nucleosynthesis_calculations}

\begin{figure}
    \includegraphics[width=\columnwidth]{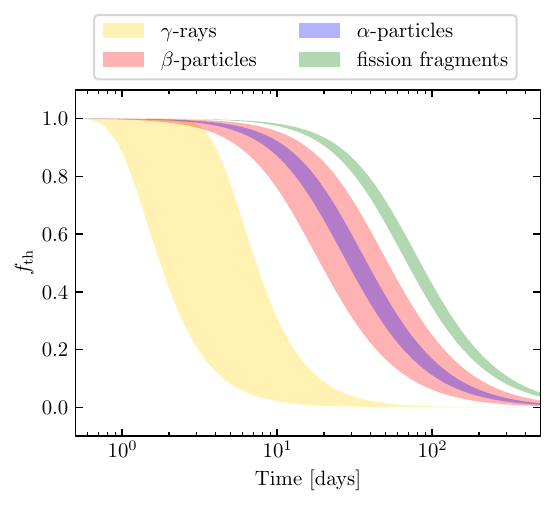}
    \caption{Thermalization efficiency of $\gamma$-rays, electrons, $\alpha$-particles and fission fragments as a function of time post-merger for typical particle injection energies, computed using \refeq{eq:therm_particles} and \refeq{eq:therm_gammas}.
    The inefficiency timescale is computed by assuming a spherical outflow with uniform density, total mass $M_{\rm ej}=0.01~M_\odot$ and characteristic velocity $v_{\rm ej}=0.1~c$.
    We consider the following ranges for the particle initial energies: $E_{0,\gamma}=0.2-1.5$ MeV, $E_{0,\beta}=0.2-1.5$ MeV, $E_{0,\alpha}=5-9$ MeV and $E_{\rm 0,ff}=100-150$ MeV \citep{Barnes:2016umi}.}
    \label{fig:therm_eff}
\end{figure}

For each merger model, we post-process the thermodynamic evolution of the ejecta using the nuclear reaction network \texttt{WinNet} \citep{Reichert:2023} in order to determine their nucleosynthesis history.
In particular, we perform network calculations on $\sim 15\,000$ tracers per merger simulation, and we investigate the effects of reducing this number on the final yields by specifically selecting a subset which is representative of the ejecta composition.
We discuss the details of this tracer selection in \refapp{app:tracer_selection}, where we note that a fraction of less than $10\%$ of the total tracers is enough to reproduce the same results without a significant loss in accuracy.

The nuclear reaction network consists of 6545 nuclear species linked with a set of temperature-dependent reaction rates imported from the JINA REACLIB library \citep{Cyburt:2010a}.
The latter includes available experimental rates, together with theoretical rates largely based on the FRDM nuclear mass model \citep{Moller:2015fba}.
Spontaneous, neutron-induced and $\beta$-delayed fissions are treated as presented in \citet{Panov:2001}, while neutrino-induced reactions are not included since they are expected to play a subdominant role at the characteristic densities and temperatures of the fluid trajectories.
Each nucleosynthesis calculation starts when the fluid temperature decreases to 7 GK, with the initial composition being determined assuming \ac{NSE}.
The tracer trajectory is then extrapolated past the simulation time by assuming adiabatic, homologous expansion.
We record until 1 Gyr the elemental abundances evolution and the radioactive energy output, which we separate into the main contributions from $\beta$-decays, $\alpha$-decays, and fissions.

In order to estimate the associated \ac{KN} emission, we treat the energy release by distinguishing between different energy carriers, i.e. $\beta$-particles, $\alpha$-particles, $\gamma$-rays, and fission fragments.
We import from the Evaluated Nuclear Data File (\href{https://www-nds.iaea.org/exfor/endf.htm}{ENDF/B-VIII.0, International Atomic Energy Agency}) the measured average energies of $\beta$- and $\alpha$-decay products for each parent nuclide. If data is not available we adopt the Q-value fractions suggested by \citet{Barnes:2016umi}, i.e. we assume that typically in $\beta$-decays $\sim20\%$ of the energy is carried by electrons, $\sim45\%$ by $\gamma$-rays and the remaining $\sim35\%$ escapes in the form of neutrinos, while in $\alpha$-decays roughly the entirety of the energy budget is retained by $\alpha$-particles.
Furthermore, we assign all the energy produced in fission processes to the kinetic energy of the fission products.

Each channel deposits energy differently within the ejecta, with an efficiency $f_{\rm th,i}(t)=\frac{\dot{Q}_{\rm dep,i}(t)}{\dot{Q}_{\rm r,i}(t)}$ gradually decreasing in time, where $\dot{Q}_{\rm r,i}(t)$ and $\dot{Q}_{\rm dep,i}(t)$ are the radioactive heating rate and the energy deposition rate for the channel $i$, respectively.
Following \citet{Barnes:2016umi}, for an outflow represented by a sphere with uniform density, the thermalization efficiency can be approximately described by
\begin{equation}\label{eq:therm_particles}
    f_{\rm th,p}(t)=\frac{{\rm ln}\left[1+2\left(\frac{t}{t_{\rm ineff,p}}\right)^2\right]}{2\left(\frac{t}{t_{\rm ineff,p}}\right)^2} \, ,
\end{equation}
\begin{equation}\label{eq:therm_gammas}
    f_{\rm th,\gamma}(t)=1-{\rm exp}\left[-\left(\frac{t}{t_{\rm ineff,\gamma}}\right)^{-2}\right] \, ,
\end{equation}
with \refeq{eq:therm_particles} and \refeq{eq:therm_gammas} valid for massive particles and $\gamma$-rays respectively.
Here $t_{\rm ineff}$ is the characteristic time at which thermalization becomes inefficient, i.e. when the expansion timescale becomes shorter than the thermalization timescale.
The latter is in general dependent on the type of decay product, on its initial energy $E_0$, and on the material density.
Within the uniform density approximation, for ejecta with mass $M_{\rm ej}$ and characteristic velocity $v_{\rm ej}$, the inefficiency time in the case of massive particles follows $t_{\rm ineff}\propto E_{\rm 0,p}^{-1/2}M_{\rm ej}^{1/2}v_{\rm ej}^{-3/2}$. In the case of $\gamma$-rays instead, $t_{\rm ineff}\propto M_{\rm ej}^{1/2}v_{\rm ej}^{-1}$, with a variation of a factor of a few depending on whether the dominant interaction is Compton scattering ($E_{\rm 0,\gamma}\gtrsim1$ MeV) or photoionization ($E_{\rm 0,\gamma}\lesssim1$ MeV) \citep{Barnes:2016umi}.
As shown in \reffig{fig:therm_eff}, for typical fluid conditions and initial energy of decay products, the thermalization of $\gamma$-rays becomes inefficient already after $\sim1$ day, while massive particles are able to efficiently deposit their energy into the medium for several days, with the more massive and slower fission fragments having the largest efficiency at each time.

\subsection{Kilonova model}
\label{sec:kilonova_model}

We couple the nucleosynthesis calculations to a semi-analytic \ac{KN} model in order to obtain synthetic \ac{KN} light curves.
We construct an anisotropic \ac{KN} scheme informed by the ejecta geometry, and we model an approximate energy transfer in order to account for the diffusion and the emission of photons.
In particular, we start from the model for spherically symmetric ejecta presented by \citet{Hotokezaka:2020} and based on the supernova model from \citet{Arnett:1982}.
While the latter describes the ejecta using a single zone approximation, the model from \citet{Hotokezaka:2020} accounts for the ejecta stratification by computing the contribution to the luminosity coming from each spherical shell with mass $m_i$, velocity $v_i$, and opacity $\kappa_i$.
The ejecta density is assumed to follow the radial profile
\begin{equation}
    \rho(t,v)=\rho_0(t)\left(\frac{v}{v_0}\right)^{-4.5} \, ,
\end{equation}
with $v$ ranging between the minimum ejecta velocity $v_0$ and the maximum velocity $v_{\rm max}$, and $\rho_0$ being defined by fixing the total ejecta mass $M_{\rm ej}$.
The model is based on the first law of thermodynamics for a radiation dominated gas, i.e.
\begin{equation}
    \frac{dE_i}{dt}=-\frac{E_i}{t}+m_i\dot{\epsilon}(t)-L_i(t) \, ,
\end{equation}
where $E_i$ is the \textit{i}-th shell internal energy, $\dot{\epsilon}$ is the specific radioactive heating rate and $L_i$ is the shell contribution to the luminosity.
The total bolometric luminosity is approximated by
\begin{equation}\label{eq:bol_lum}
    L(t)=\sum_i{\rm erfc}\left(\sqrt{\frac{t_{{\rm diff},i}}{2t}}\right)\frac{E_i}{{\rm min}(t_{{\rm diff},i},t)+\frac{v_it}{c}} \, ,
\end{equation}
with $t_{{\rm diff},i}=\tau_{\gamma,i}v_itc^{-1}$ the photon diffusion timescale and $\tau_{\gamma,i}$ the photon optical depth, defined as
\begin{equation}
    \tau_{\gamma,i}(t)=\int_{v_i}^{v_{\rm max}}\kappa(v)\rho(v)tdv \, .
\end{equation}
We characterize the ejecta opacity radial profile $\kappa(v)$ by employing a step function with an average inner opacity value $\kappa_{\rm in}$, an average outer value $\kappa_{\rm out}$ in general different from the first, and the step $v_{\rm step}$ located at the approximate interface between the dynamical ejecta and the disk ejecta.
The complementary error function term in \refeq{eq:bol_lum} approximately describes the fraction of energy that escapes from each shell.
For a diffusion timescale which is greater than the expansion timescale, photons are essentially trapped in the medium and most of the radiation energy does not contribute to the instantaneous luminosity.
On the other hand, when the diffusion timescale is shorter than the expansion timescale, radiation escapes almost freely and the contributing fraction is $\sim1$.

Following \citet{Martin:2015hxa,Perego:2017wtu,Barbieri:2019sjc,Barbieri:2019kli,Camilletti:2022jms,Ricigliano:2024}, we generalize the spherical model to the anisotropic case by assuming axisymmetry with respect to the system rotational axis and by discretizing the ejecta along the polar angle $\theta$ direction in 15 slices uniform in $\theta$.
To each angular bin we assign the total mass of the fluid elements ejected at the corresponding latitude.
The ejection angle is evaluated using the element last tracked displacement, i.e. we assume to follow the ejecta long enough to resolve their asymptotic direction.
Concurrently, we also associate to the angular bin a mass-weighted average velocity together with a minimum and a maximum velocity value which are represented by the $1\%$ and the $99\%$ mass-weighted percentiles of the bin tracers velocity distribution respectively, in order to exclude unrelevant outliers.
The velocity of each tracer is evaluated using the asymptotic definition $v_\infty=\sqrt{1-\left(\frac{h_\infty}{hu_t}\right)^2}$, following, e.g., \citet{Fujibayashi:2020dvr}.
Finally, we compute mass-weighted average opacities for the inner and the outer part of the angular bin, on the basis of the initial electron fraction $Y_{\rm e}$.
The opacity parametrization is imported from \citet{Tanaka:2019iqp}, which perform systematic atomic structure calculations for all the elements between $26<Z<88$, and compute a representative Planck grey opacity for each composition in a set characterized by the value of initial $Y_{\rm e}$.
Although approximate, these estimates return a reasonable representation of the overall opacity in the ejecta with typical densities $\rho\sim10^{-13}~{\rm g~cm^{-3}}$ and temperatures $T\lesssim20000$ K, where thermal photons diffusion is dominated by bound-bound transitions in mildly-ionized r-process elements.

The latitudinal bin mass is isotropized with a factor $4\pi/\Omega$, where $\Omega$ is the bin angular size, and it is injected together with the bin characteristic velocities, opacities and the thermalized heating rate from the nucleosynthesis calculations in the spherical \ac{KN} model, in order to compute the bin isotropized contribution to the total bolometric luminosity.
The latter is then obtained by summing all the angular contributions rescaled back to the original bin size.

We describe the radiation escaping from the outflow as an approximate black-body with a surface represented by the ejecta photosphere.
At each latitude, the photosphere position $r_{\rm ph}$ in time is tracked by imposing the condition $\tau_\gamma(r_{\rm ph})\sim1$ to the photon optical depth, and the photospheric effective temperature $T_{\rm ph}$ is computed using the Stefan-Boltzmann law.
In a luminosity peak timescale, the photosphere receeds inwards in the outflow, and its effective temperature decreases from $\sim10000$ K to a few 1000 K.
We impose a minimum to $T_{\rm ph}$ as the photosphere starts to follow the electron-ion recombination in the late expanding ejecta \citep{Barnes:2013wka}, and we parametrize this temperature floor with the local composition.
We choose as representative bracketing values $T_{\rm floor}^{\rm La}=2000$ K for Lanthanides-polluted compositions with a low initial electron fraction $Y_{\rm e}\lesssim0.2$ and $T_{\rm floor}^{\rm Ni}=4000$ K for Lanthanides-poor compositions with $Y_{\rm e}\gtrsim0.3$.

We derive the AB magnitudes at a given frequency $\nu$ as
\begin{equation}
    m_{\rm AB,\nu}(t)=-2.5~{\rm log}_{10}(f_{\nu}(t))-48.6 \, ,
\end{equation}
where $f_{\nu}(t)$ is the observed flux, i.e.
\begin{equation}
    f_{\nu}(t)=\frac{1}{4\pi D_L^2}\int p(\theta,\theta_{\rm view})\frac{L_{\rm iso}(\theta,t)}{\sigma_{\rm SB} T_{\rm ph}^4(\theta,t)}B_{\nu}(T_{\rm ph}(\theta,t))d\theta \, .
\end{equation}
Here $D_L$ is the luminosity distance from the source, which we fix to 40 Mpc, i.e. the value inferred for the gravitational wave signal GW170817, while $L_{\rm iso}(\theta,t)$ is the isotropized bolometric luminosity contribution from the angular bin at $\theta$, $B_\nu(T)$ is the Planck distribution, $\sigma_{\rm SB}$ is the Stefan-Boltzmann constant and $p(\theta,\theta_{\rm view})$ is the flux projection factor in the direction of the observer at a viewing angle $\theta_{\rm view}$.

%% ==============================================================
%%
%%                           RESULTS
%%
%% ==============================================================

\section{Results}
\label{sec:results}

\subsection{Ejecta properties}
\label{sec:ejecta_properties}

\begin{figure}
    \includegraphics[width=.8\columnwidth]{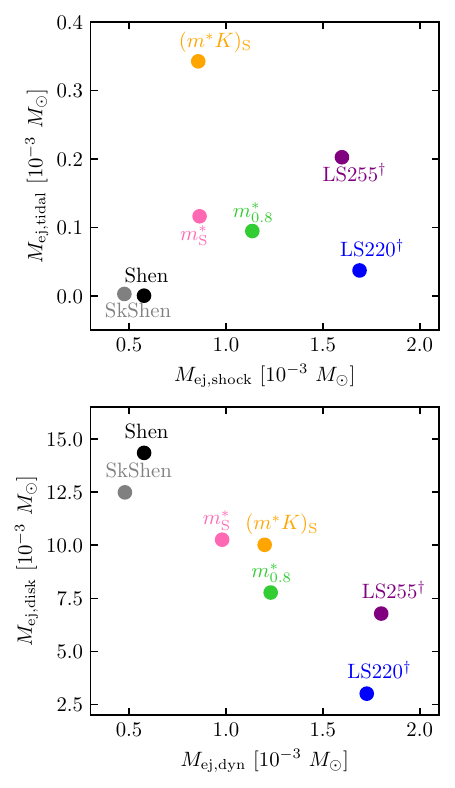}
    \caption{Total masses of each ejecta component for the different \ac{EOS} models: tidal ejecta versus shock-heated ejecta (top panel) and dynamical ejecta versus disk ejecta (bottom panel).
    The masses are extracted at the end of the simulations and are classified according to the procedure described in \refsec{sec:ejecta_extraction}.}
    \label{fig:ejecta_masses}
\end{figure}

\begin{figure*}
    \includegraphics[width=.9\textwidth]{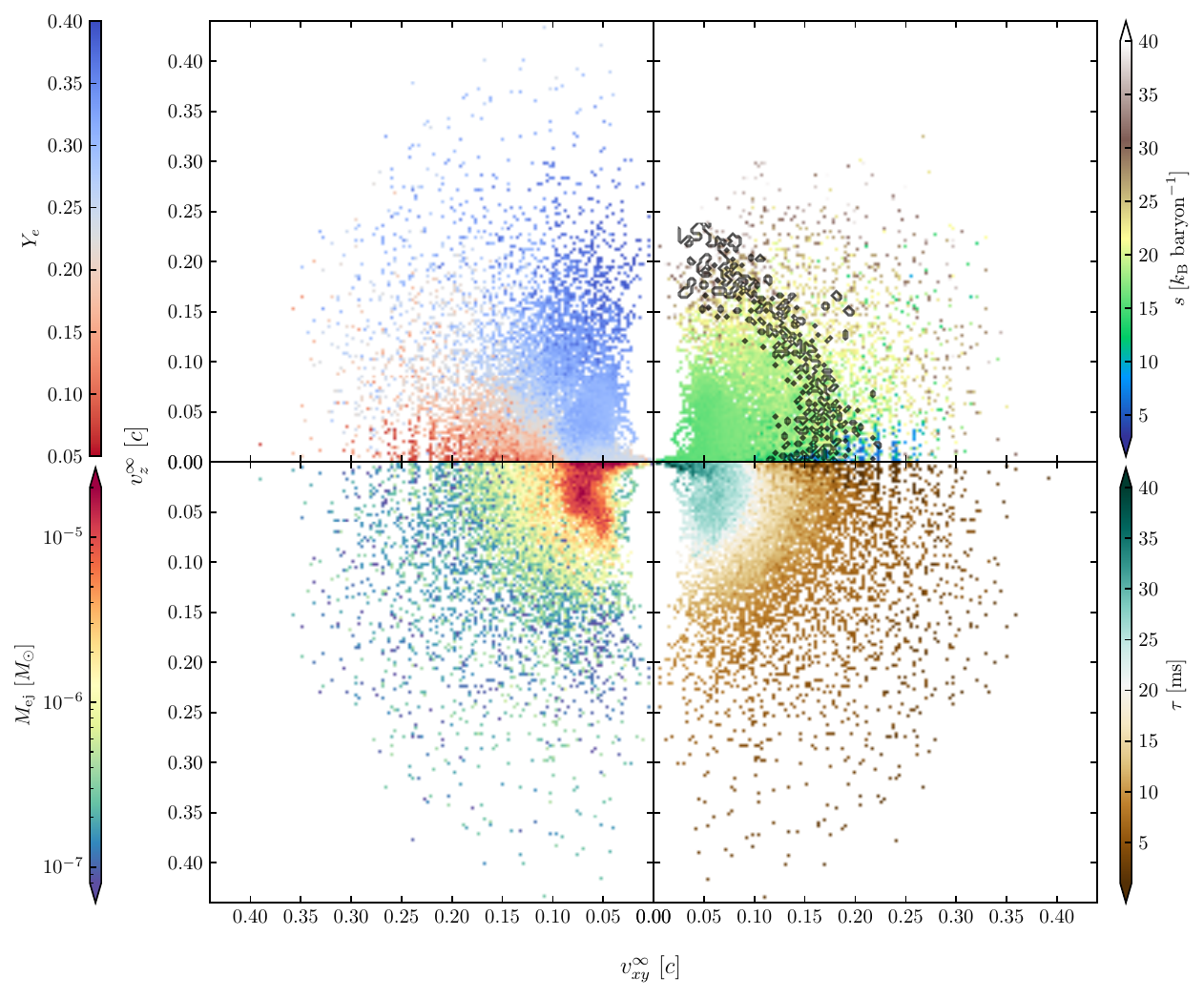}
    \caption{Representation of tracers on a velocity space grid for the model \msS.
    Points in the grid are colour-coded relating to the mass-weighted average electron fraction (top left panel), specific entropy (top right panel) and expansion timescale (bottom right panel), evaluated at a tracer temperature of $\sim5.8$ GK.
    The bottom left panel shows the tracer's mass distribution.
    The dark grey interface in the top right panel represents the approximative boundary between the slower disk ejecta and the faster dynamical ejecta, as computed using the geodesic criterion.}
    \label{fig:section}
\end{figure*}

\begin{figure*}
    \includegraphics[width=\textwidth]{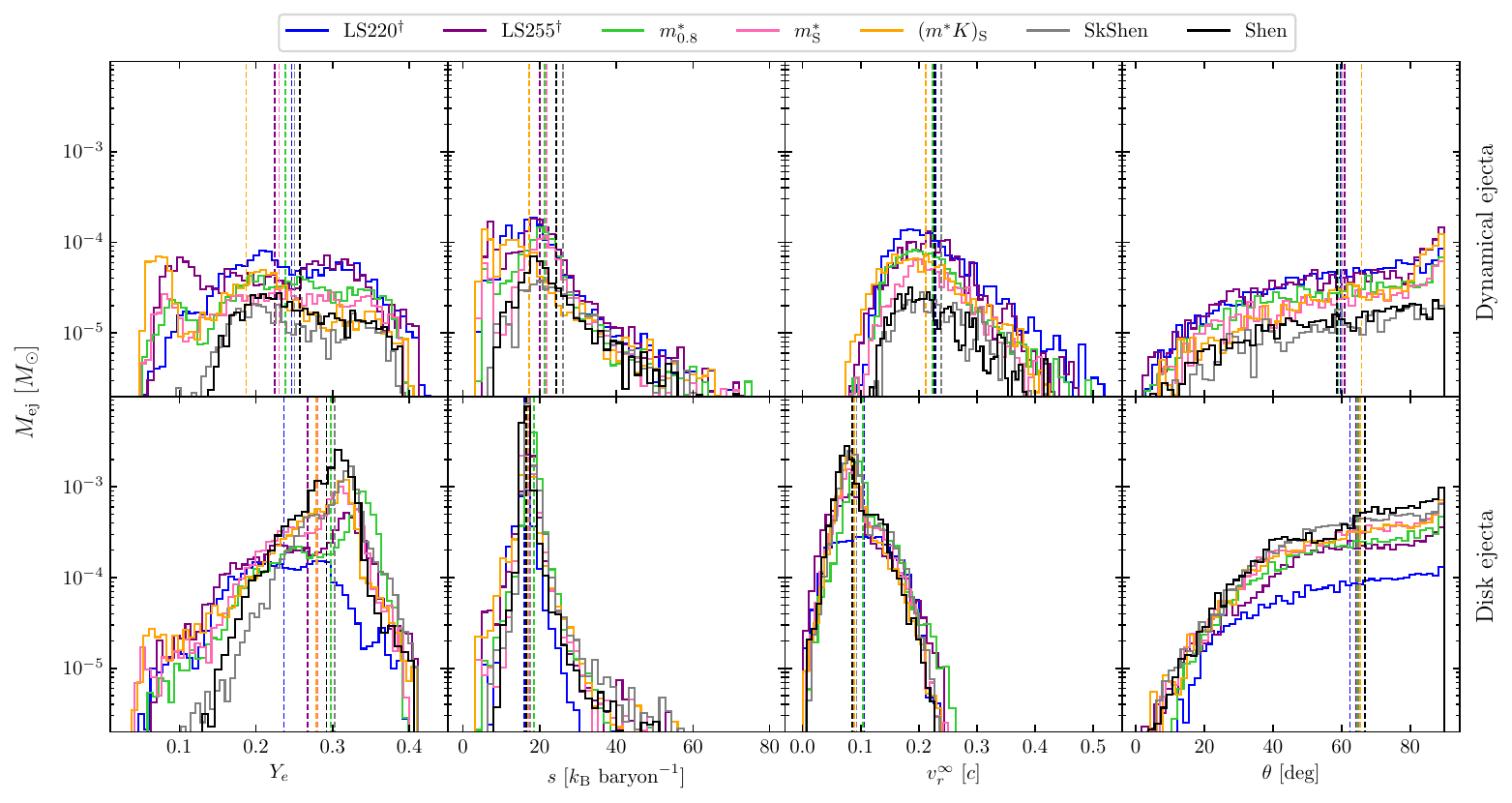}
    \caption{Ejecta mass histograms as a function of the electron fraction $Y_e$, the specific entropy $s$, the asymptotic radial velocity $v^\infty_r$ and the polar angle of ejection $\theta$, evaluated at a tracer temperature of $\sim5.8$ GK, for the different \ac{EOS} models, separately for the dynamical ejecta (top panels) and for the early disk ejecta (bottom panels).
    Vertical dashed lines represent the mass-weighted average values of each quantity.}
    \label{fig:hist}
\end{figure*}

As investigated in many previous works \citep[see, e.g., ][]{Bauswein:2010dn, Hotokezaka:2012ze, Sekiguchi:2016bjd, Radice:2018pdn, Nedora:2021eojb}, the \ac{EOS} strongly influence the evolution of the binary system in its merging and post-merging phase, impacting the amount of dynamical and disk ejecta.
Figure \ref{fig:ejecta_masses} shows the masses of the ejecta components (as defined in \refsec{sec:ejecta_extraction}) for all the models considered in this work.
The dynamical ejecta are split into tidal and shock-heated ejecta (top panel of \reffig{fig:ejecta_masses}).
\citet{Jacobi:2023} found that the mass of the shock-heated component increases for higher pressures at saturation density, while the mass of the tidal component grows with the slope of the pressure as a function of density, in the same density regime.
In particular, the models that differ only in the incompressibility (\LSm and \LSh as well as \msS and \msKS) exhibit roughly the same amount of shock-heated ejecta but vary considerably in the amount of tidal ejecta.
On the other hand, in the models that differ only in the effective mass (\LSm, \msm and \msS), a smaller variation in the amount of tidal ejecta is found, while the amount of shock-heated ejecta varies considerably.
The models SkShen and Shen do not exhibit any ejected material with $Y_e\lesssim0.1$.
In these models, the tidally ejected matter is reached by shocks from the merger interface very quickly, and all of its content is thus reprocessed towards higher electron fractions.

The amount of secular ejecta is highly dependent on the final fate of the central remnant \citep[see, e.g.,][]{Fujibayashi:2022ftg}.
Mergers that result in a long-lived massive neutron star exhibit a larger accretion disk (since otherwise roughly half of the disk is accreted upon collapse), leading to more ejecta from the accretion disk system.
Furthermore, the ejection of the matter is enhanced by the reabsorption of neutrinos emitted by the remnant \citep{Perego:2014, Just:2014fka, Radice:2018} and spiral-wave winds driven by its oscillations \citep{Nedora:2019jhl, Nedora:2020pak}.

In the model \LSm, a hyper-massive neutron star forms after the merger and survives for $\sim \unit[8]{ms}$.
Consequently, in such a case the disk ejecta mass is very small.
All other models considered in this work do not form a \ac{BH} in the simulated time ($\sim40-50$ ms).
For them, the registered mass of the disk ejecta roughly increases with the stiffness of the underlying \ac{EOS}, and it is significantly larger than that of the dynamical ejecta.
However, we note that this trend is far from solid, since there is still matter ejected at the end of each simulation, and thus the mass values are only lower limits.
In addition, it is expected that in all models a sizable fraction of the accretion disk will also be ejected on timescales of seconds due to alternative mechanisms, such as viscous heating \citep[see, e.g., ][]{Hotokezaka:2012ze,Fernandez:2018kax,Fujibayashi:2020dvr,Fahlman:2022jkh}.

Despite the different masses of the ejecta components, many general features are consistent across all models.
Figure \ref{fig:section} shows the distribution of the electron fraction $Y_e$, specific entropy $s$, mass $M_{\rm ej}$, and expansion timescale  $\tau=r(v_r^\infty)^{-1}$ (all of which evaluated at $T\sim5.8$ GK) versus their equatorial and polar asymptotic velocity for the model \msS.
The dynamical ejection channel operates within the first $\sim10$ ms post-merger, unbinding neutron-rich ($Y_e\lesssim0.1$), low-entropy ($s\lesssim10~k_{\rm B}~{\rm baryon}^{-1}$) material, predominantly along the equatorial plane, with velocities $\sim0.15-0.35~c$.
The shock-heated ejecta are launched more isotropically with higher entropies up to $\sim40~k_{\rm B}~{\rm baryon}^{-1}$, traveling at velocities up to $\sim0.45~c$.
This material is heated significantly and thus positron captures increase the electron fraction to $0.1\lesssim Y_e\lesssim0.4$.
The electron fraction gradually increases with the latitude, as a result of the lower density and higher entropy in the polar region.
In addition, despite having velocities similar to those of the tidal component, the temperature of the shocked matter reaches \unit[5.8]{GK} only at later times because it is initially hotter.
Its expansion timescale is of the order of $\sim \unit[10]{ms}$, compared to that of the tidal ejecta, which is $\lesssim \unit[5]{ms}$.

Outflows from the disk emerge after $\sim10$ ms post-merger, with lower velocities $\lesssim0.15~c$ and correspondingly longer expansion timescales up to $\sim \unit[40]{ms}$.
This component shows fairly constant entropies $s\sim20~k_{\rm B}~{\rm baryon}^{-1}$ and intermediate values of the electron fraction $0.15\lesssim Y_e\lesssim0.4$.
The disk ejecta is launched in all directions, with its mass density decreasing towards higher latitudes.
However, it does not fully cover the polar area ($\theta\lesssim20^\circ$), because it is mainly driven by the remnant density waves traveling on the equatorial plane.
In addition, neutrino absorption mostly impacts the outer exposed layers of the disk, causing the matter ejection to be more efficient at intermediate latitudes.
The transition between dynamical ejecta and disk ejecta is depicted in \reffig{fig:section} as a dark grey, slightly prolate, thick interface where material from different ejection mechanisms is present.

The mass-weighted histograms of the ejecta properties for the different models are shown in \reffig{fig:hist}.
The electron fraction and the entropy distributions of the dynamical ejecta have a systematic double-peak structure due to the tidal and the shock-heated component.
The tidal component forms the low $Y_e$, low entropy peak, which is present in all models except for the SkShen and Shen \acp{EOS}, and it is enhanced in the models \LSh and \msKS (see the discussion above).
Instead, the shock-heated component is responsible for the higher $Y_e$ plateau and the second peak at larger entropies.

Regarding the disk ejecta, the electron fraction distribution typically peaks at $Y_e \approx 0.3$.
This peak is caused by the absorption of neutrinos in the ejecta above the disk (see \citet{Jacobi:2023} for more information), and constitutes the largest fraction of the disk ejecta.
The same matter also forms the peak in the velocity histogram at $v^\infty_r\approx0.08~c$.
Such component is missing in the \LSm model, since the collapse of the remnant into \ac{BH} leads both to a suppression of the ejection and to a reduction in the neutrino reprocessing to which the already present ejecta is exposed.
Notably, the specific position of the high $Y_e$ peak, as well as the overall average $Y_e$ in the ejecta, does not exhibit a clear correlation with the amount of matter expelled from the disk.
It is instead clearer a dependency of the peak position on the disk latitudinal extension, which partially shields the ejecta from neutrino irradiation.
Thus, for the model \msm, where the disk is considerably thinner than all the other non-collapsing models, the electron fraction shows the highest peak value, at $Y_e\sim0.33$.

\subsection{Nucleosynthesis yields}
\label{sec:nucleosynthesis_yields}

\begin{figure}
    \includegraphics[width=\columnwidth]{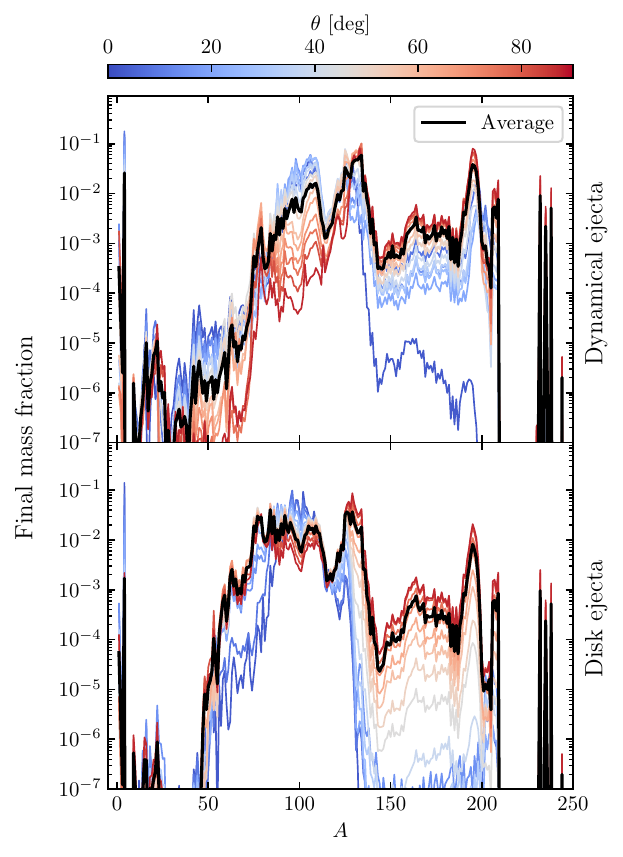}
    \caption{Final abundance patterns in the dynamical ejecta (top panel) and the early disk ejecta (bottom panel), relative to different polar angles $\theta$ for the model \msS.
    Each pattern is obtained through a mass-weighted average of the yields of all the tracers within the corresponding angular bin centered in $\theta$.
    The overall mass-weighted average of each ejecta component is reported in black.}
    \label{fig:finab_ang}
\end{figure}

\begin{figure}
    \includegraphics[width=\columnwidth]{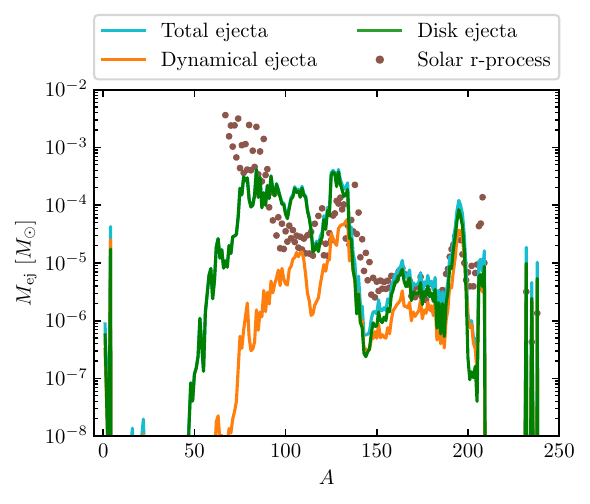}
    \caption{Final elemental mass distribution for the model \msS, along with the separate contributions from the dynamical ejecta and the disk ejecta.
    The solar r-process abundances \citep{lodders2009AbundancesElementsSolar} normalized to the model abundance of Os ($A=190$) are shown as reference.}
    \label{fig:finab_comp}
\end{figure}

\begin{figure}
    \includegraphics[width=\columnwidth]{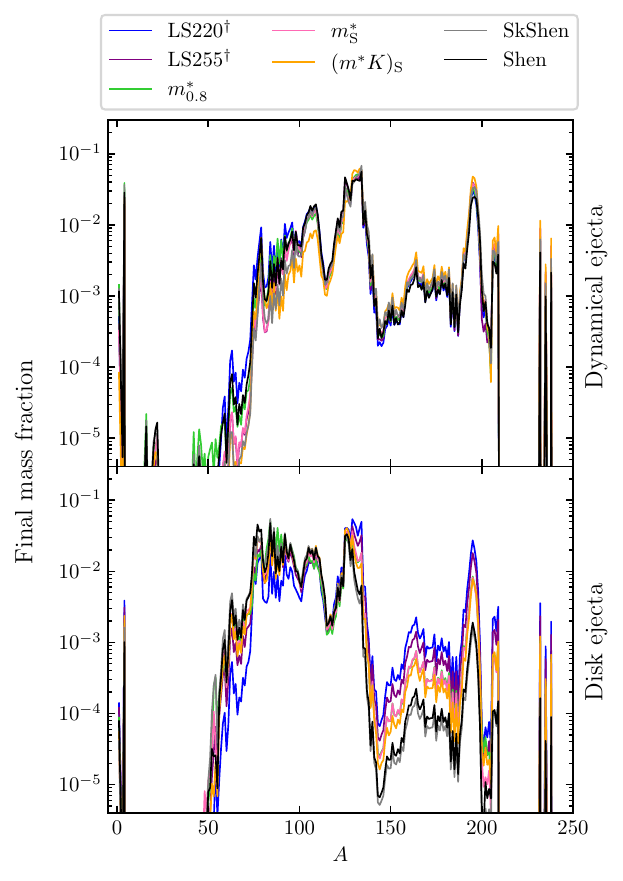}
    \caption{Final abundance pattern in the dynamical ejecta (top panel) and in the disk ejecta (bottom panel) for each \ac{EOS} model.
    Each yield is obtained through a mass-weighted average of the yields of all the tracers belonging to the specific ejecta component from the corresponding model.}
    \label{fig:finab}
\end{figure}

\begin{figure*}
    \includegraphics[width=\textwidth]{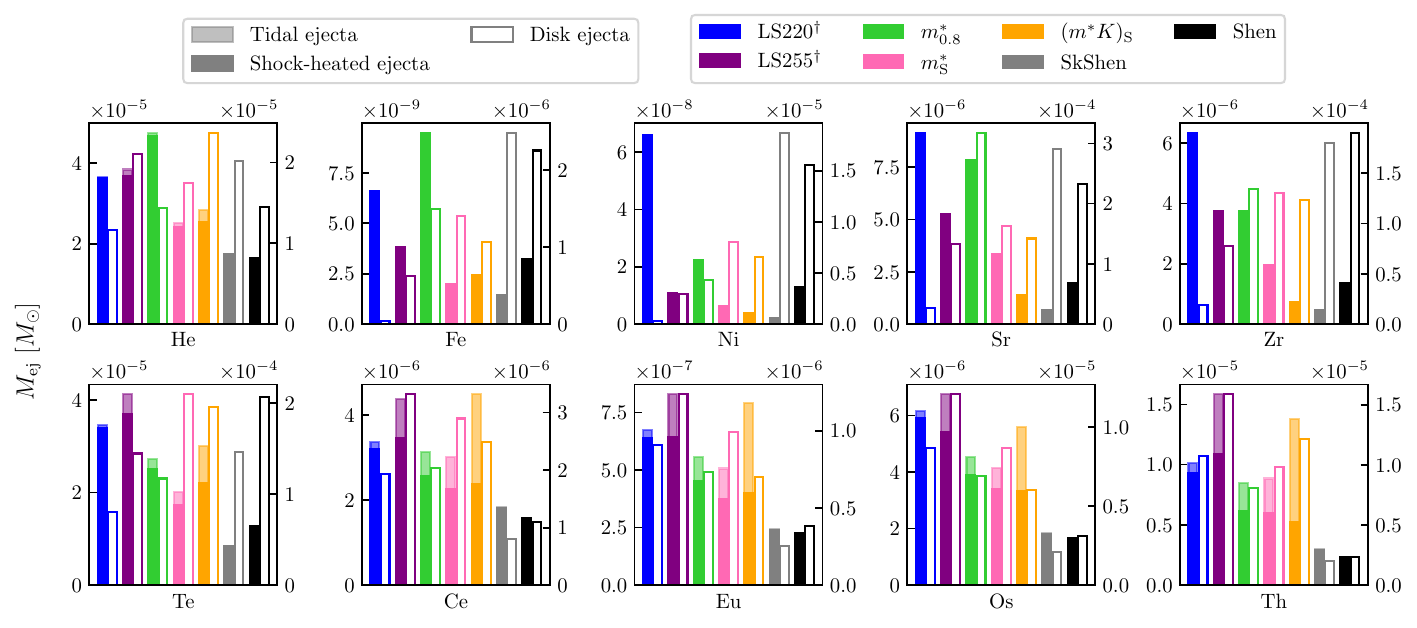}
    \caption{Final mass of a set of elements in the dynamical ejecta (left axis) and the disk ejecta (right axis) for each \ac{EOS} model.
    For the dynamical ejecta, the contributions from the tidal and the shock-heated component are shown.}
    \label{fig:finab_elem}
\end{figure*}

\begin{figure}
    \includegraphics[width=\columnwidth]{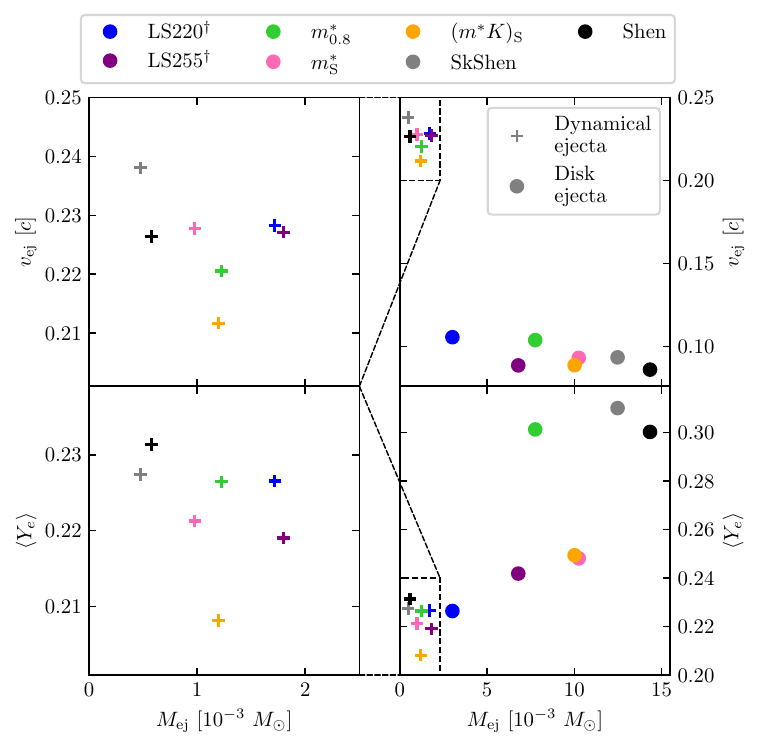}
    \caption{Dynamical ejecta and disk ejecta for each \ac{EOS} model, represented in the space of total mass vs average velocity (top panel) and vs average electron fraction (bottom panel)}
    \label{fig:mass_vel_ye}
\end{figure}

We use the output of the nucleosynthesis calculations described in \refsec{sec:nucleosynthesis_calculations} to characterize the composition in the ejecta.
We focus on the final abundances, which are representative of the ejecta composition as it eventually contributes to the galaxy enrichment.

\reffig{fig:finab_ang} shows the average abundance pattern as a function of the polar angle $\theta$ for one representative model, i.e. \msS.
Both in the dynamical ejecta and in the early disk ejecta, the composition is strongly dependent on the angular direction, due to the latitudinal anisotropy in the initial electron fraction, entropy, and velocity of the ejecta.
As seen in \refsec{sec:ejecta_properties}, the material ejected has a lower electron fraction close to the equatorial plane and a higher electron fraction in the polar region.
The distribution is wider in the case of the dynamical ejecta, as it comprehends also the very low $Y_e\lesssim0.1$ material.
This feature is directly reflected in the angular patterns, whereas at high latitudes the composition is generally dominated by lighter elements, while at low latitudes a stronger r-process occurs, due to the larger availability of free neutrons.
Regarding the dynamical component, its equatorial pattern features both second and third r-process peak elements to constitute more than $\sim20\%$ of the local material, together with a suppression of the first r-process peak.
In contrast, the polar pattern shows an inverted scenario, with the first peak and the lighter elements enhanced, the second peak partially decreased and the third peak almost absent.
However, while along the latitudinal direction there is a large variability in the abundance of light elements, we find that a considerable amount of heavy elements is still produced down to angles $\theta\sim20^\circ$, despite the local higher average electron fraction.
This has to be ascribed to the longer expansion timescale of the shock-heated component ($\tau\gtrsim10$ ms), which allows for higher entropy r-process \citep[see, e.g.,][]{Lippuner:2015gwa}.
On the other hand, in the early disk ejecta, the angular scatter in the first r-process peak is substantially reduced to one order of magnitude at most, while elements beyond the second peak show a more gradual variability, with the third peak still moderately present at the equator but progressively disappearing close to the pole.
This behaviour is expected for mild ejecta velocities $\sim0.1~c$, intermediate entropies $\sim20~{\rm k_B~baryon^{-1}}$, and for electron fractions lying close to the intermediate value $Y_e\sim0.25$, denoting the strong sensitivity of the composition in this parameter region \citep[similar results for the disk outflow composition were found in, e.g.,][]{Curtis:2023}.
In this regard, we note that the presence of a moderate third peak for low latitudes in the disk ejecta is not particularly dependent on the separation criterion for the ejecta components that we adopted.
For example, by increasing the extraction radius for the $u_t$ evaluation to 400 km, the amount of low-$Y_e$ material counting as dynamical ejecta instead of disk ejecta increases, due to the additional kinetic energy gained over its thermal energy, but it is found to be non-determinant in this context.

By the final simulated time, the overall composition is already mostly dominated by the disk ejecta, as evidenced in \reffig{fig:finab_comp}, which compares the total ejecta composition with the contributions from the dynamical ejecta and the early disk ejecta for the representative model \msS.
The only exception is constituted by the model \LSm, for which the final amounts of disk ejecta and dynamical ejecta are comparable.
For all the models with a stable remnant, the dynamical component contribution to the final abundances is completely irrelevant for species with $50\lesssim A\lesssim 130$.
For $A\gtrsim130$ we find only a minor impact, which is expected to be reduced further, as ejection from the disk continues in time.
However the additional ejecta are also expected to have higher electron fractions, across the different phases in the disk evolution, thus roughly freezing the yields of elements heavier than $A\gtrsim130$ \citep[as found in, e.g.,][]{Martin:2015hxa, Magistrelli:2024}.
Interestingly, \reffig{fig:finab_comp} also shows that in all models the light species with $A\lesssim50$, together with the heavier species with $200\lesssim A\lesssim205$, are still dominated by the contribution of the dynamical ejecta, as a result of the much greater specific yields of these elements returned in such component.
This is due to the larger spread in the thermodynamic conditions of the dynamical material.
For example, $\alpha$ particles are produced both in higher $Y_e$ and $s$ environments (polar shock-heated outflow), at the charged-particle reaction freeze-out, as well as in lower $Y_e$ and $s$ regions (tidal ejecta), also thanks to a contribution from the decay of heavy elements \citep[see, e.g., ][for a recent work on this subject]{Perego:2020evn}.

We also note that in the final patterns the ratio between second and third r-process peak elements is comparable to the one of the solar r-process \citep{lodders2009AbundancesElementsSolar}.
Rare earth elements are decently reproduced, while the second peak only roughly agrees with the solar one, being narrower for all models. 
Elements with $130\lesssim A\lesssim155$ are systematically underproduced, which can most-likely be attributed to the current uncertainties in the nuclear inputs of the model \citep{Horowitz:2019, Cowan:2019pkx}.
In addition, there is no agreement for elements with $A\lesssim115$, however, in such region, viscous disk ejecta from the same neutron star merger event or ejecta from other astrophysical sources are likely to contribute as well.

In general, when comparing the final abundances across the different models, we find that the dynamical ejecta host a robust r-process pattern in all cases, varying at most of a factor of a few for elements heavier than $A\sim110$ (see \reffig{fig:finab}).
Greater variability is found for elements with $A\lesssim110$, with up to one order of magnitude scatter for the first r-process peak.
The production of rare earth and third peak elements is slightly enhanced moving from the Shen to the \msKS model, for two reasons:
\begin{itemize}
    \item the gradually increasing amount of tidal material expelled, characterized by very low electron fractions $Y_e\leq0.1$ at equatorial angles of ejection (see top panels in \reffig{fig:hist}), systematically creates such species;
    \item these species are also produced in the increasing shock-heated component, by matter with $0.1\leq Y_e\lesssim0.3$ and a sufficiently broad range of entropies and expansion timescales.
\end{itemize}
\reffig{fig:finab_elem} shows the total ejected mass of a series of elements, divided into the contributions from the tidal, shock-heated and disk ejecta components. 
Within the dynamical ejecta, the contribution to the heavy elements production from the shock-heated component is dominant in most cases.
Even for lanthanides such as Ce or Eu, the tidal contribution constitutes at most only half of the total yield.
Therefore, for these elements a decrease in the amount of tidal ejecta is partially compensated by the increase in shock-heated ejecta.
The first r-process peak region and lighter elements instead exhibit a trivial enhancement for greater amounts of shock-heated ejecta, as in the \LSm, \LSh and \msm models. 
However, the precise yields are substantially sensitive to the specific conditions found in this ejecta component.
In particular, across different models we find variations in the ratio of elements already within the same group, such as between Fe and Ni or Sr and Zr.
Therefore, for the dynamical ejecta, the amount of heavy elements produced in a strong r-process as well as the amount of lighter elements, hardly show any evident correlation with the deeper model parameters that characterize the original nuclear matter.
This statement could be revised in absence of the heterogeneous shock-heated ejecta, as it would happen in a \ac{BHNS} merger or a prompt \ac{BH} formation case, in which only the tidal ejecta could dynamically emerge.

Regarding the disk ejecta yields, the \msm, SkShen and Shen models typically return the highest amount of light elements up to the first r-process peak, while second and third peak are produced more in models such as \LSm and \LSh (see both \reffig{fig:finab} and \reffig{fig:finab_elem}).
Notably, however, the average initial $Y_e$ is not a precise proxy for the final abundances, as the latter depend on the details of the $Y_e$ distribution.
For example, while the relative amount of iron group elements roughly follows the average $Y_e$, first peak elements are remarkably more sensitive to the position of the maximum in the $Y_e$ distribution.
Furthermore, as visible in \reffig{fig:mass_vel_ye}, the model \msm has among the highest values of average $Y_e$ and position of the maximum in the $Y_e$ distribution, but it still produces lanthanides yields which are comparable to the model \msKS.

In turn, the $Y_e$ distribution evolves with the total amount of ejecta, since the initial conditions in the expelled material vary with time.
Thus, the production of heavy elements across different models is less robust in the disk ejecta compared to the dynamical ejecta.
On one side, this can be connected to the relative subdominance of a very low $Y_e$ contribution within this channel, and on the other side to the lower velocity and entropy conditions, which can help building up most of the species already at high temperatures, depending on the specific $Y_e$.
In general, the relative abundance of heavy elements only roughly scales inversely with the amount of disk ejecta, which are constituted by mixed-$Y_e$ material.
It could be reasonable to extend this scaling relation to the general stiffness of the \ac{EOS}, but more data is necessary in order to verify this possibility.

\subsection{Kilonova light curves}
\label{sec:kilonova_lightcurves}

\begin{figure}
    \includegraphics[width=\columnwidth]{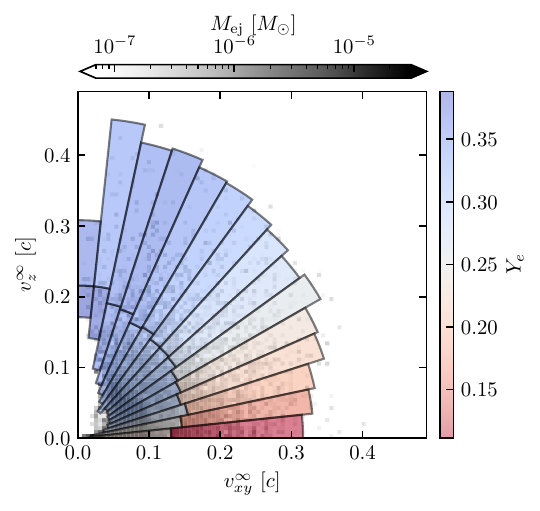}
    \caption{Scheme of the ejecta from the model \msS, as it is read by the semi-analytic \ac{KN} model.
    Each sector represents a different angular bin, where the model is run on a ray-by-ray basis, with its boundaries in the radial direction being defined by the mass-weighted $1\%$ and $99\%$ percentiles of the tracers distribution.
    Each sector is divided into two zones, characterized by a different average gray opacity, which is parametrized with the local value of electron fraction, following \citet{Tanaka:2019iqp}.
    The division between the two bin zones is obtained by averaging the position of the interface between the dynamical ejecta and the disk ejecta components.
    The ejecta mass histogram is shown as reference in gray in the background.}
    \label{fig:kn_scheme}
\end{figure}

\begin{figure}
    \includegraphics[width=\columnwidth]{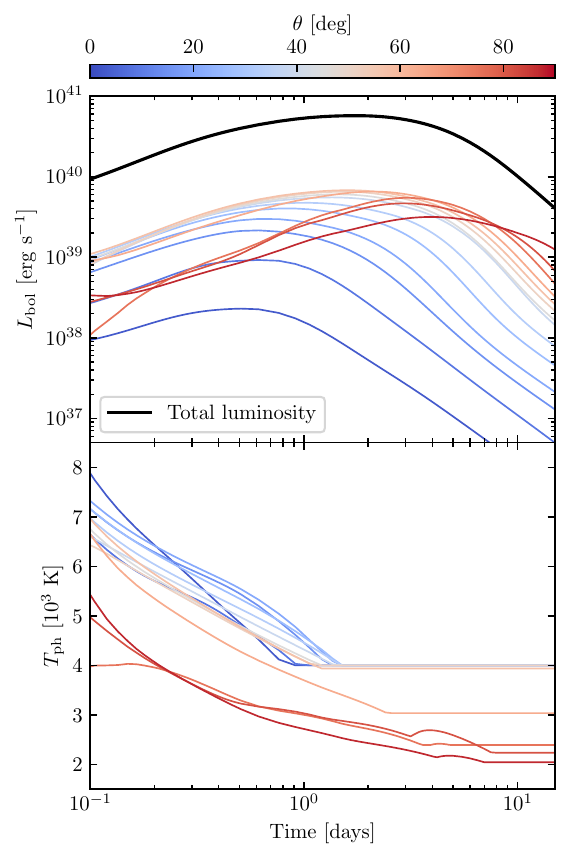}
    \caption{Total bolometric luminosity for the model \msS, along with the contribution coming from different angular bins (top panel).
    Effective photospheric temperature evolution as a function of the ejecta angle for the same model (bottom panel).
    A composition-dependent floor value is applied to the temperature in order to account for recombination in the late expanding ejecta.}
    \label{fig:lumbol_tph}
\end{figure}

\begin{figure}
    \includegraphics[width=\columnwidth]{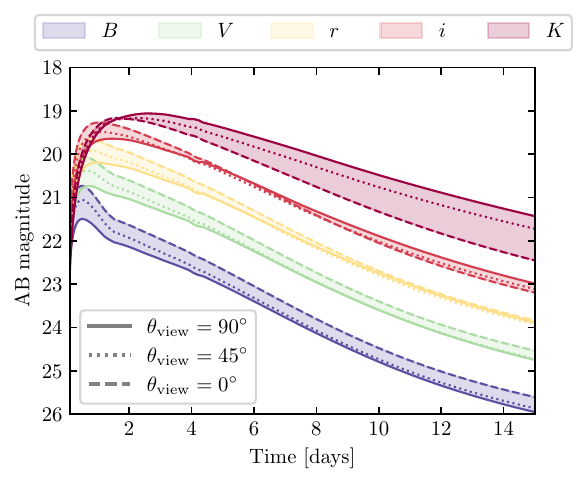}
    \caption{Broad-band light curves in various optical/IR filters for the model \msS.
    The dependency on the observer direction is shown, with distinct line styles indicating different viewing angles, varying from the polar direction (dashed lines) to the equatorial one (solid lines).}
    \label{fig:mags}
\end{figure}

\begin{figure*}
    \includegraphics[width=\textwidth]{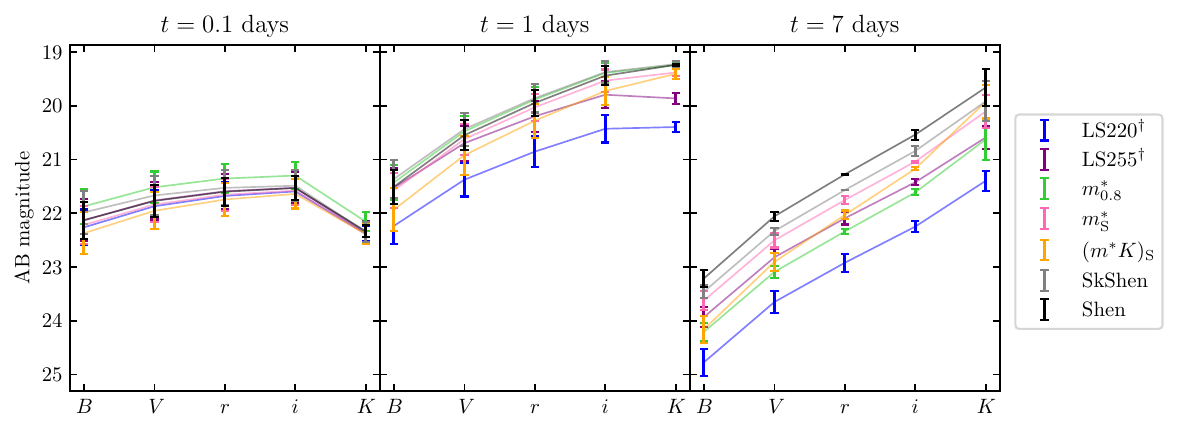}
    \caption{Color magnitudes at 0.1, 1 and 7 days in different filters for all the \ac{EOS} models.
    The vertical bars indicate the variability related to the observer direction.}
    \label{fig:mags_eos}
\end{figure*}

Given the ejecta properties obtained from the extracted sets of tracers (\refsec{sec:ejecta_properties}) and the corresponding nucleosynthesis (\refsec{sec:nucleosynthesis_yields}), we now investigate the subsequent thermal \ac{KN} emission.
We compute the bolometric and color light curves within $\sim10$ days post-merger using our anisotropic framework (\refsec{sec:kilonova_model}), and we investigate the potential impact of the \ac{EOS} on this transient.
The ejecta scheme depicted in \reffig{fig:kn_scheme} is useful to understand which ejecta features are carried over to the \ac{KN} model. By construction, the latter simplifies the ejecta structure in order to solve analytically the radiative transfer problem.
All models display an approximately prolate ejecta shape, with material in the polar region being typically $10\%~c$ faster than the equatorial direction.
However, the density significantly decreases at higher latitudes, and, as a result, fewer and lower mass tracers populate the polar angular bins.
Excluding outliers, this population shows a much narrower velocity distribution.
Concurrently, the computed average opacities reflect the compositional variety of each component:
\begin{itemize}
    \item a wider opacity range, varying from $\sim2~{\rm cm^2~g^{-1}}$ at the pole to $\sim30~{\rm cm^2~g^{-1}}$ at the equator, is found in the external, dynamically driven ejecta layers;
    \item a shorter opacity range, with values between $\sim2~{\rm cm^2~g^{-1}}$ and $\sim10~{\rm cm^2~g^{-1}}$, is observed in the internal regions originating from the disk.
\end{itemize}
As a consequence, the radial opacity profile decreases with larger radii at high latitudes, and it instead increases significantly with the radius at lower latitudes, potentially inducing the lanthanides curtain effect \citep[see, e.g.,][]{Kasen:2014toa,Wollaeger:2017ahm}.
According to the latter, bluer radiation can be shielded from an hypothetical edge-on observer ($\theta_{\rm view}=90^\circ$) by the lanthanide-rich material in the outermost layers of the ejecta.

With these features being roughly common to all the considered models, several aspects of the resulting emission are found persistent, regardless of the \ac{EOS} employed.
First of all, the bolometric luminosity peaks between 1 and 3 days, when the ejecta has expanded enough for its average optical depth to become comparable to the mean photon diffusion path.
At that time the ejecta radiates with a rate of a few $10^{40}~{\rm erg~s^{-1}}$, catching up with the instantaneuous energy deposition from the radioactive heating from r-process material.
However, we note that the specific values of peak magnitude and timescale are sensitive to the total amount and type of ejecta produced in the merger event.
As investigated further in \refsec{sec:extension}, we expect the luminosity peak to shift to later times and to be brighter, once the disk ejecta contribution is fully accounted for, because of the additional fuel provided to power the \ac{KN} emission.
As visible in \reffig{fig:lumbol_tph}, within the peak timescale, the total luminosity of the transient is dominated by the contribution of the ejecta regions at intermediate latitudes.
In these regions, the balance between the local higher density and the lower material opacity is optimal.
In fact, on one hand the fraction of energy emerging from the strictly polar region at its peak is at least one order of magnitude lower than from the equator, due to the very low amount of mass ejected in such direction.
On the other hand, the equatorial contribution is characterized by a higher opacity as a result of the higher concentration of lanthanides, causing the radiation to be trapped for longer times and to be released at lower temperatures.
Red photons eventually escape after a few days, when the overall emission has passed its peak, and become the dominant contribution.
Notably, the considerable diffusion of a large variety of heavy elements throughout the ejecta and the subsequent dominance of their $\beta$-decay has the effect of suppressing any peculiar feature in the average radioactive heating rates, within the time frame of the \ac{KN} photospheric phase.
Such features would include the decay of particularly abundant lighter nuclear species, potentially visible in a weak r-process scenario, or the signature of fissioning nuclei, which do not emerge within the first $\sim15$ days post-merger.

As described in \refsec{sec:kilonova_model}, in order to compute the color evolution, we approximate the emission continuum with a standard black-body spectrum emerging from the ejecta photosphere and parametrized by an effective photospheric temperature $T_{\rm ph}$.
This approximation was found to be reasonable within a few days post-merger in the case of AT2017gfo, since in this time frame the presence of specific spectral features has a negligible impact on the overall evolution of the color curves \citep{Smartt:2017,Watson:2019xjv}.
At higher latitudes, where the local photosphere receeds inwards faster, the photospheric temperatures are instantaneously higher than in proximity of the equator, and the resulting emission is bluer.
In the considered time interval for the \ac{KN}, average temperatures in the ejecta are found to be lower than $\sim10000$ K.
This allows us to employ the grey opacities suggested by \citet{Tanaka:2019iqp} and computed for low-level (I-IV) ionized atoms, as can be typically found at those temperatures and densities.
On the contrary, a careful exploration of the early time emission would require more extensive opacity calculations, including higher ionization stages in significantly hotter material, as investigated by \citet{Banerjee:2020myd,Banerjee:2022doa,Banerjee:2023gye}.

The color evolution of the thermal transient for the representative \msS model is shown in \reffig{fig:mags}.
The impact of the ejecta mass and opacity structure is directly reflected in the relatively early dominance of the redder bands, which rapidly take over the bluer ones already within $\sim1$ day post-merger.
The emission in these bands is further boosted at later times by progressively looking at the ejecta from a more edge-on configuration, due to the maximization of the projected area corresponding to the more opaque ejecta region.
On the contrary, bluer bands are generally multiple orders of magnitude less luminous, with a relatively weaker enhancement for a face-on view ($\theta_{\rm view}=0^\circ$).
The main contribution to the latter emerges from a less dense environment with, however, an average opacity which never decreases down to typical values of Fe-like elements $\sim0.1~{\rm cm^2~g^{-1}}$.
Nevertheless, once again, we recall that the role of the blue emission must be validated against the presence of additional moderate to high $Y_e$ material, potentially ejected on longer timescales from the disk (see \refsec{sec:extension}).

In terms of light curves, the most impacting effect of varying the \ac{EOS} model is the overall amount of material expelled.
As shown in \reffig{fig:mags_eos}, the emission brightness roughly follows the total ejecta mass, with a progressively brighter transient moving from the \LSm to the Shen model. 
The latter gains on \LSm more than one magnitude at peak time in all bands, while delaying the peak timescale of $\sim2$ days.
For all the \ac{EOS} models, we observe a progressive broadening in the IR band variability associated with the observer viewing angle.
This feature derives from the faster fading of the polar contribution with respect to the equatorial one.
Concurrently, the trend is inverted for the bluer bands, with a gradual isotropization of the emission, due to the disappearance of the initially enhanced blue polar contribution.
The effect is slower for the \LSm model, as a result of the greater general anisotropy produced by its dynamical ejecta coupled with the smaller disk ejecta.
Among the different models, \msm is the one generating the bluer and faster evolving transient, due to the small relative amount of lanthanides.
The Shen and SkShen have similar compositions, but they shine longer thanks to the additional disk ejecta mass.
On the other hand, with respect to similar models, \msKS misses part of the early blue emission because the lanthanides curtain extends to relevantly higher latitudes.

In the context of an observed emission, the primary correlation between its brightness and the ejecta mass can be used as a constraint on the general \ac{EOS} stiffness, as already pointed out by, e.g., \citet{Bauswein:2013yna,Radice:2018pdn}.
Such a dependence becomes particularly determinant when the collapse of the central remnant is involved, as in the case of the \LSm model.
In principle, combinations of nuclear matter properties concurring to the increase in the \ac{EOS} softness could be disfavoured by the potential detection of a too bright \ac{KN}, although a similar effect could also be achieved by an asymmetric enough initial binary.
However, even in such case, it would not be clear how to constrain these nuclear matter properties individually, breaking the degeneracy that appears in the process.

\subsection{Kilonova model extension}
\label{sec:extension}

\begin{figure}
    \includegraphics[width=\columnwidth]{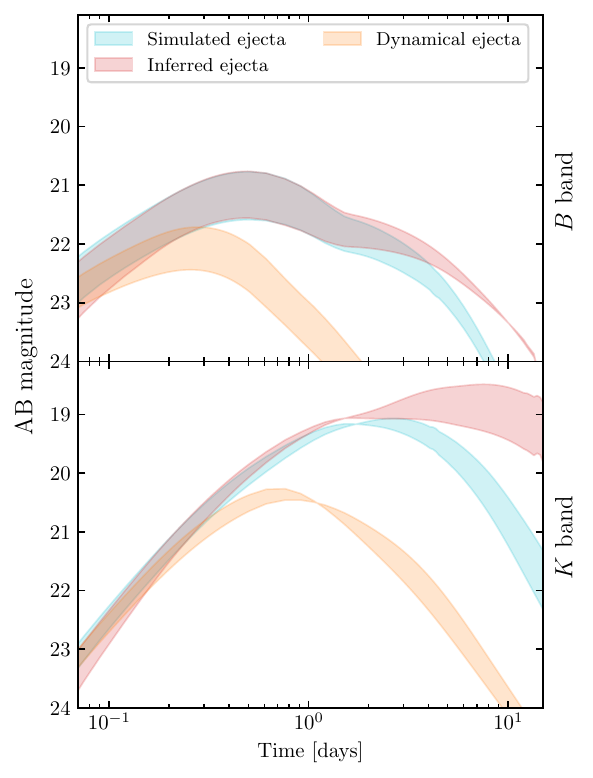}
    \caption{Broad-band light curves in the $B$ and $K$ filters for the model \msS, obtained by considering the ejecta from the simulation, in comparison to the result of including an additional secular ejecta component.
    The latter is modeled starting from the last portion of disk ejecta simulated and assuming an ejected mass of $20\%$ of the disk mass.
    For visual reference, the light curves resulting from ignoring all the ejecta components past the dynamical ejecta are also shown.
    The spread of the curves indicates to the variability related to the observer direction.}
    \label{fig:mags_ext}
\end{figure}

The final elemental yields in \refsec{sec:nucleosynthesis_yields} and the \ac{KN} light curves in \refsec{sec:kilonova_lightcurves} are sensitive to the ending time of our simulations.
In order to determine whether this dependency impacts our results regarding the \ac{EOS} effects, we model a hypothetical secular ejecta, stemming from the disk on longer timescales.
We assume that roughly an additional $20\%$ of the disk mass is ejected through secular mechanisms, as generally found in simulations of remnant disks accreting onto the central object \citep[see, e.g., ][]{Hotokezaka:2012ze,Just:2014fka,Fernandez:2018kax,Fujibayashi:2020dvr,Just:2021cls,Fahlman:2022jkh}.
Our assumption is supported by the fact that in all models the disk mass is found to be saturated by the end of the simulation.
This results in a secular ejecta mass between $\sim0.1~M_\odot$ and $\sim0.06~M_\odot$, depending on the model.
For such ejecta component, we adopt the properties of the last $\sim10\%$ of disk mass ejected, i.e. we extract the velocity, entropy, electron fraction, and angular distributions of matter expelled approximately after $\sim40$ ms post-merger.
We thus obtain a secular ejecta where velocities and entropies closely follow the ones of the earlier disk ejecta, with an average velocity of $\sim0.07~c$ and average entropy of $\sim18~k_{\rm B}~{\rm baryon}^{-1}$.
However, the ejection is mostly concentrated at angles $\gtrsim50^\circ$, and the electron fraction is distributed only between $0.25\lesssim Y_e\lesssim0.35$, with its peak shifted systematically to $Y_e\gtrsim0.3$.
Therefore, the resulting composition does not include elements past the second r-process peak.

We show in \reffig{fig:mags_ext} the \ac{KN} light curves for the representative \msS model, obtained by taking the secular ejecta into account, in comparison to our previous results using the original ejecta.
The additional late outflow enriches the slow, optically thick bulk of the ejecta with radioactive material exhibiting mild opacity values $\sim5~{\rm cm^2~g^{-1}}$.
As a result, most of the injected energy is radiated away only at later times and lower frequencies, once the ejecta has expanded and cooled enough.
The emission peak in the IR is thus shifted to $\sim4$ days post-merger, with an increase in the peak luminosity of $\sim1$ magnitude.
The impact of the secular ejecta is less evident for times $\lesssim1$ day, when the emission is mostly dominated by the outer layers.
In fact, the bluer emission component is affected only in its decaying phase, well after its peak.
If we assume that this toy model broadly reproduces a plausible secular ejection in its main features, then this result confirms that in our models the early, blue phase of the \ac{KN} is mainly dependent on the properties of the dynamical ejecta and of the early disk ejecta.
However, we find less straightforward to identify a time frame where the light curves are completely determined by the sole dynamical ejecta component.
\reffig{fig:mags_ext} also shows the light curves obtained by ignoring all the ejecta components expelled after the dynamical ejecta.
In such a case, the emission is affected already during the first hours after merger, with the largest variations for the bluer bands.
The latter are found to be dimmer by around $\sim1$ magnitude at 7 hours post-merger, with a faster fading, due to the lack of the contribution from lower opacity disk ejecta material.
In light of this result, as pointed out in \refsec{sec:kilonova_lightcurves}, early \ac{KN} observations could help in constraining the general stiffness of the \ac{EOS}, which is likely correlated to the amount of matter ejected in the first tens of milliseconds post-merger.
Nevertheless, according to our findings, it would be difficult to extract information useful to constrain specific nuclear matter properties, such as the incompressibility or the effective nucleon mass.
Such an inference would be possible only in the hypothesis that there is a time frame in which the \ac{KN} is uniquely determined by the dynamical ejecta contribution.
This scenario would naturally exclude the formation of a massive accretion disk, and would favor a fast collapse of the remnant in order to suppress any neutrino-driven wind \citep[see, e.g.][]{Perego:2014,Radice:2018,Kiuchi:2019lls}.
In any case, assuming an available early enough detection, the benchmark modeling would still require more detailed physics treatment, in primis on the opacity side.

%% ==============================================================
%%
%%                        CONCLUSION
%%
%% ==============================================================

\section{Conclusion}
\label{sec:conclusion}

In this work, we reprised the discussion on the \ac{EOS} of nuclear matter, in relation to \ac{BNS} merger events.
We considered a set of \ac{BNS} merger GRHD simulations employing different \ac{EOS} models \citep{Yasin:2020a,Jacobi:2023}, and we investigated the associated nucleosynthesis and \ac{KN}.
For this purpose, we used the widely adopted tracer extraction method, in order to inform a nuclear reaction network, coupled with a ray-by-ray semi-analytic \ac{KN} model.

We focused on the variation of the incompressibility parameter and of the effective nucleon mass, which affect the pressure and thermal evolution inside the merger remnant, leading to different amount of material ejected.
For example, the softest models, \LSl and \LSm, provoke a prompt and delayed collapse of the remnant respectively, and no or little ejecta is produced.
On the other hand, for higher incompressibilities and lower effective masses, the overall ejection in the first tens of milliseconds post-merger is generally boosted, due to the longer-surviving oscillations of the remnant.
Since these ejecta stem from the remnant disk with a broad range of initial electron fractions centered around $Y_e\sim0.25$, the resulting nucleosynthesis drives an r-process with variable strength.
Third peak elements are consistently found in the regions around the binary orbital plane, and the contribution from the disk outflows is comparable to the one of the dynamical ejecta.
Variations in the final abundances across different \ac{EOS} models are contained within one order of magnitude.
Within the parameter range considered, these abundances exhibit a sensitivity in primis on the details of the initial $Y_e$ distribution, which is not accurately captured by only one representative value, such as the average $Y_e$.
A stronger r-process pattern is found for smaller mass ejections, which roughly result from models with a softer \ac{EOS}.
However, a link to the nuclear matter parameters emerges only when considering the dynamical ejecta alone.
In such context, the incompressibility and the effective nucleon mass show a connection with the amount of tidal and shock-heated ejecta respectively.
In particular, the shock-heated component can significantly contribute to the production of elements as heavy as Eu or Pt.
Therefore, a scenario where this channel is suppressed by a prompt \ac{BH} presence could isolate the role of the incompressibility in producing such elements through purely tidal ejections.

Extending the discussion to the \ac{KN} emission, light curves in the first dozen days post-merger are found relatively featureless, because of the widespread lanthanides-rich composition blending the radioactive energy deposition.
The \ac{EOS} directly impacts the emission peak magnitude, whereas stiffer \acp{EOS} generally produce more ejecta and brighter transients.
Concurrently, the composition distribution affects the transient evolution, with the fastest evolving emission produced by the \msm model, where a smaller amount of material ejected combines with a lower lanthanides pollution.
\ac{KN} light curves are the result of a complex chain of processes, and a large degeneracy emerges already when determining the sourcing ejecta properties.
In light of the emission response to the \ac{EOS} variation, tracing it back to the nuclear matter properties appears even more uncertain, and constrains can be reasonably put only on the overall \ac{EOS} stiffness.
Furthermore, light curves and final abundance patterns, as computed from the simulation output, are affected by an additional uncertainty, caused by the short simulated physical time frame.
We expect potentially important differences, once the full ejection history is taken into account.
Outflows from the disk after the simulated few tens of ms are expected to enrich the ejecta with higher $Y_e$ material, and the light curves are altered in a way to substantially delay the emission peak.
However, it appears clear that the disk ejecta play at least a partial role in producing the early blue \ac{KN} component as well as the later red one.
Therefore, we find difficult to isolate features in the \ac{KN} emission which can be traced back to the sole dynamical ejecta, and, by extension, to details of the original nuclear \ac{EOS}.

We interpret this result as a warning towards the study of the nuclear \ac{EOS} using merger yields and \ac{KN} light curves.
This approach could still hold valuable information when combined to \ac{GW} analysis, in extended, systematic studies, in order to face the large degeneracy of the problem.

\section*{Acknowledgements}

The authors are grateful to A. Schwenk, F. M. Guercilena, K. Hotokezaka, M. Reichert, and S. Huth, for the valuable contributions and feedbacks in relation to the production of this work.
This work is supported by the Deutsche Forschungsgemeinschaft (DFG, German Research Foundation, Project ID 279384907, SFB 1245), and by the State of Hesse within the Research Cluster ELEMENTS (Project ID 500/10.006).

\section*{Data availability}

The data concerning this work will be shared on reasonable request to the authors.

%% ===============================================================
\bibliographystyle{mnras}
\bibliography{refs,local}

\begin{thebibliography}{}
\makeatletter
\relax
\def\mn@urlcharsother{\let\do\@makeother \do\$\do\&\do\#\do\^\do\_\do\%\do\~}
\def\mn@doi{\begingroup\mn@urlcharsother \@ifnextchar [ {\mn@doi@} {\mn@doi@[]}}
\def\mn@doi@[#1]#2{\def\@tempa{#1}\ifx\@tempa\@empty \href {http://dx.doi.org/#2} {doi:#2}\else \href {http://dx.doi.org/#2} {#1}\fi \endgroup}
\def\mn@eprint#1#2{\mn@eprint@#1:#2::\@nil}
\def\mn@eprint@arXiv#1{\href {http://arxiv.org/abs/#1} {{\tt arXiv:#1}}}
\def\mn@eprint@dblp#1{\href {http://dblp.uni-trier.de/rec/bibtex/#1.xml} {dblp:#1}}
\def\mn@eprint@#1:#2:#3:#4\@nil{\def\@tempa {#1}\def\@tempb {#2}\def\@tempc {#3}\ifx \@tempc \@empty \let \@tempc \@tempb \let \@tempb \@tempa \fi \ifx \@tempb \@empty \def\@tempb {arXiv}\fi \@ifundefined {mn@eprint@\@tempb}{\@tempb:\@tempc}{\expandafter \expandafter \csname mn@eprint@\@tempb\endcsname \expandafter{\@tempc}}}

\bibitem[\protect\citeauthoryear{Abbott et~al.}{Abbott et~al.}{2017a}]{TheLIGOScientific:2017qsa}
Abbott B.~P.,  et~al., 2017a, \mn@doi [Phys. Rev. Lett.] {10.1103/PhysRevLett.119.161101}, 119, 161101

\bibitem[\protect\citeauthoryear{Abbott et~al.}{Abbott et~al.}{2017b}]{GBM:2017lvd}
Abbott B.~P.,  et~al., 2017b, \mn@doi [Astrophys. J.] {10.3847/2041-8213/aa91c9}, 848, L12

\bibitem[\protect\citeauthoryear{Abbott et~al.}{Abbott et~al.}{2018}]{Abbott:2018exr}
Abbott B.~P.,  et~al., 2018, \mn@doi [Phys. Rev. Lett.] {10.1103/PhysRevLett.121.161101}, 121, 161101

\bibitem[\protect\citeauthoryear{Abbott et~al.}{Abbott et~al.}{2019}]{Abbott:2018wiz}
Abbott B.~P.,  et~al., 2019, \mn@doi [Phys. Rev.] {10.1103/PhysRevX.9.011001}, X9, 011001

\bibitem[\protect\citeauthoryear{Annala, Gorda, Kurkela  \& Vuorinen}{Annala et~al.}{2018}]{Annala:2017llu}
Annala E.,  Gorda T.,  Kurkela A.,   Vuorinen A.,  2018, \mn@doi [Phys. Rev. Lett.] {10.1103/PhysRevLett.120.172703}, 120, 172703

\bibitem[\protect\citeauthoryear{{Arnett}}{{Arnett}}{1982}]{Arnett:1982}
{Arnett} W.~D.,  1982, \mn@doi [Astrophys. J.] {10.1086/159681}, \href {https://ui.adsabs.harvard.edu/abs/1982ApJ...253..785A} {253, 785}

\bibitem[\protect\citeauthoryear{Banerjee, Tanaka, Kawaguchi, Kato  \& Gaigalas}{Banerjee et~al.}{2020}]{Banerjee:2020myd}
Banerjee S.,  Tanaka M.,  Kawaguchi K.,  Kato D.,   Gaigalas G.,  2020, \mn@doi [Astrophys. J.] {10.3847/1538-4357/abae61}, 901, 29

\bibitem[\protect\citeauthoryear{Banerjee, Tanaka, Kato, Gaigalas, Kawaguchi  \& Domoto}{Banerjee et~al.}{2022}]{Banerjee:2022doa}
Banerjee S.,  Tanaka M.,  Kato D.,  Gaigalas G.,  Kawaguchi K.,   Domoto N.,  2022, \mn@doi [Astrophys. J.] {10.3847/1538-4357/ac7565}, 934, 117

\bibitem[\protect\citeauthoryear{Banerjee, Tanaka, Kato  \& Gaigalas}{Banerjee et~al.}{2023}]{Banerjee:2023gye}
Banerjee S.,  Tanaka M.,  Kato D.,   Gaigalas G.,  2023, preprint (ArXiv:2304.05810)

\bibitem[\protect\citeauthoryear{Barbieri, Salafia, Perego, Colpi  \& Ghirlanda}{Barbieri et~al.}{2019}]{Barbieri:2019sjc}
Barbieri C.,  Salafia O.~S.,  Perego A.,  Colpi M.,   Ghirlanda G.,  2019, \mn@doi [Astron. Astrophys.] {10.1051/0004-6361/201935443}, 625, A152

\bibitem[\protect\citeauthoryear{Barbieri, Salafia, Perego, Colpi  \& Ghirlanda}{Barbieri et~al.}{2020}]{Barbieri:2019kli}
Barbieri C.,  Salafia O.~S.,  Perego A.,  Colpi M.,   Ghirlanda G.,  2020, \mn@doi [Eur. Phys. J.] {10.1140/epja/s10050-019-00013-x}, A56, 8

\bibitem[\protect\citeauthoryear{Barnes \& Kasen}{Barnes \& Kasen}{2013}]{Barnes:2013wka}
Barnes J.,  Kasen D.,  2013, \mn@doi [Astrophys. J.] {10.1088/0004-637X/775/1/18}, 775, 18

\bibitem[\protect\citeauthoryear{Barnes, Kasen, Wu  \& Martinez-Pinedo}{Barnes et~al.}{2016}]{Barnes:2016umi}
Barnes J.,  Kasen D.,  Wu M.-R.,   Martinez-Pinedo G.,  2016, \mn@doi [Astrophys. J.] {10.3847/0004-637X/829/2/110}, 829, 110

\bibitem[\protect\citeauthoryear{Bauswein, Janka  \& Oechslin}{Bauswein et~al.}{2010}]{Bauswein:2010dn}
Bauswein A.,  Janka H.-T.,   Oechslin R.,  2010, \mn@doi [Phys. Rev.] {10.1103/PhysRevD.82.084043}, D82, 084043

\bibitem[\protect\citeauthoryear{Bauswein, Janka, Hebeler  \& Schwenk}{Bauswein et~al.}{2012}]{Bauswein:2012ya}
Bauswein A.,  Janka H.,  Hebeler K.,   Schwenk A.,  2012, \mn@doi [Phys. Rev.] {10.1103/PhysRevD.86.063001}, D86, 063001

\bibitem[\protect\citeauthoryear{Bauswein, Goriely  \& Janka}{Bauswein et~al.}{2013}]{Bauswein:2013yna}
Bauswein A.,  Goriely S.,   Janka H.-T.,  2013, \mn@doi [Astrophys. J.] {10.1088/0004-637X/773/1/78}, 773, 78

\bibitem[\protect\citeauthoryear{Bauswein, Just, Janka  \& Stergioulas}{Bauswein et~al.}{2017}]{Bauswein:2017vtn}
Bauswein A.,  Just O.,  Janka H.-T.,   Stergioulas N.,  2017, \mn@doi [Astrophys. J.] {10.3847/2041-8213/aa9994}, 850, L34

\bibitem[\protect\citeauthoryear{Bernuzzi, Dietrich  \& Nagar}{Bernuzzi et~al.}{2015}]{Bernuzzi:2015rla}
Bernuzzi S.,  Dietrich T.,   Nagar A.,  2015, \mn@doi [Phys. Rev. Lett.] {10.1103/PhysRevLett.115.091101}, 115, 091101

\bibitem[\protect\citeauthoryear{Blacker, Kochankovski, Bauswein, Ramos  \& Tolos}{Blacker et~al.}{2024}]{Blacker:2024}
Blacker S.,  Kochankovski H.,  Bauswein A.,  Ramos A.,   Tolos L.,  2024, \mn@doi [Phys. Rev. D] {10.1103/PhysRevD.109.043015}, 109, 043015

\bibitem[\protect\citeauthoryear{Bliss, Witt, Arcones, Montes  \& Pereira}{Bliss et~al.}{2018}]{Bliss:2018nhk}
Bliss J.,  Witt M.,  Arcones A.,  Montes F.,   Pereira J.,  2018, \mn@doi [Astrophys. J.] {10.3847/1538-4357/aaadbe}, 855, 135

\bibitem[\protect\citeauthoryear{Bombaci \& Logoteta}{Bombaci \& Logoteta}{2018}]{Bombaci:2018ksa}
Bombaci I.,  Logoteta D.,  2018, \mn@doi [Astron. Astrophys.] {10.1051/0004-6361/201731604}, 609, A128

\bibitem[\protect\citeauthoryear{Bovard \& Rezzolla}{Bovard \& Rezzolla}{2017}]{Bovard:2017}
Bovard L.,  Rezzolla L.,  2017, \mn@doi [Class. Quant. Grav.] {10.1088/1361-6382/aa8d98}, 34

\bibitem[\protect\citeauthoryear{Bovard, Martin, Guercilena, Arcones, Rezzolla  \& Korobkin}{Bovard et~al.}{2017}]{Bovard:2017mvn}
Bovard L.,  Martin D.,  Guercilena F.,  Arcones A.,  Rezzolla L.,   Korobkin O.,  2017, \mn@doi [Phys. Rev.] {10.1103/PhysRevD.96.124005}, D96, 124005

\bibitem[\protect\citeauthoryear{Bulla}{Bulla}{2019}]{Bulla:2019muo}
Bulla M.,  2019, \mn@doi [Mon. Not. Roy. Astron. Soc.] {10.1093/mnras/stz2495}, 489, 5037

\bibitem[\protect\citeauthoryear{Bulla}{Bulla}{2023}]{Bulla:2023}
Bulla M.,  2023, \mn@doi [Mon. Not. Roy. Astron. Soc.] {10.1093/mnras/stad232}, 520, 2558

\bibitem[\protect\citeauthoryear{Camilletti et~al.,}{Camilletti et~al.}{2022}]{Camilletti:2022jms}
Camilletti A.,  et~al., 2022, \mn@doi [Mon. Not. Roy. Astron. Soc.] {10.1093/mnras/stac2333}, 516, 4760

\bibitem[\protect\citeauthoryear{Combi \& Siegel}{Combi \& Siegel}{2023}]{Combi:2023}
Combi L.,  Siegel D.~M.,  2023, \mn@doi [Astrophys. J.] {10.3847/1538-4357/acac29}, 944, 28

\bibitem[\protect\citeauthoryear{Coughlin et~al.}{Coughlin et~al.}{2018}]{Coughlin:2018miv}
Coughlin M.~W.,  et~al., 2018, \mn@doi [Mon. Not. Roy. Astron. Soc.] {10.1093/mnras/sty2174}, 480, 3871

\bibitem[\protect\citeauthoryear{Cowan, Sneden, Lawler, Aprahamian, Wiescher, Langanke, Mart\'\i{}nez-Pinedo  \& Thielemann}{Cowan et~al.}{2021}]{Cowan:2019pkx}
Cowan J.~J.,  Sneden C.,  Lawler J.~E.,  Aprahamian A.,  Wiescher M.,  Langanke K.,  Mart\'\i{}nez-Pinedo G.,   Thielemann F.-K.,  2021, \mn@doi [Rev. Mod. Phys.] {10.1103/RevModPhys.93.015002}, 93, 15002

\bibitem[\protect\citeauthoryear{Curtis, Miller, Fröhlich, Sprouse, Lloyd-Ronning  \& Mumpower}{Curtis et~al.}{2023}]{Curtis:2023}
Curtis S.,  Miller J.~M.,  Fröhlich C.,  Sprouse T.,  Lloyd-Ronning N.,   Mumpower M.,  2023, \mn@doi [Astrophys. J. Lett.] {10.3847/2041-8213/acba16}, 945, L13

\bibitem[\protect\citeauthoryear{{Cyburt} et~al.,}{{Cyburt} et~al.}{2010}]{Cyburt:2010a}
{Cyburt} R.~H.,  et~al., 2010, \mn@doi [Astrophys.\ J.\ Suppl.] {10.1088/0067-0049/189/1/240}, \href {https://ui.adsabs.harvard.edu/abs/2010ApJS..189..240C} {189, 240}

\bibitem[\protect\citeauthoryear{Côté et~al.}{Côté et~al.}{2019}]{Cote:2018qku}
Côté B.,  et~al., 2019, \mn@doi [Astrophys. J.] {10.3847/1538-4357/ab10db}, 875, 106

\bibitem[\protect\citeauthoryear{Dietrich, Coughlin, Pang, Bulla, Heinzel, Issa, Tews  \& Antier}{Dietrich et~al.}{2020}]{Dietrich:2020efo}
Dietrich T.,  Coughlin M.~W.,  Pang P. T.~H.,  Bulla M.,  Heinzel J.,  Issa L.,  Tews I.,   Antier S.,  2020, \mn@doi [Science] {10.1126/science.abb4317}, 370, 1450

\bibitem[\protect\citeauthoryear{Drischler, Hebeler  \& Schwenk}{Drischler et~al.}{2016}]{Drischler:2016}
Drischler C.,  Hebeler K.,   Schwenk A.,  2016, \mn@doi [Phys. Rev. C] {10.1103/PhysRevC.93.054314}, 93, 54314

\bibitem[\protect\citeauthoryear{Drischler, Hebeler  \& Schwenk}{Drischler et~al.}{2019}]{Drischler:2019}
Drischler C.,  Hebeler K.,   Schwenk A.,  2019, \mn@doi [Phys. Rev. Lett.] {10.1103/PhysRevLett.122.042501}, 122, 042501

\bibitem[\protect\citeauthoryear{Eichler, Livio, Piran  \& Schramm}{Eichler et~al.}{1989}]{Eichler:1989ve}
Eichler D.,  Livio M.,  Piran T.,   Schramm D.~N.,  1989, \mn@doi [Nature] {10.1038/340126a0}, 340, 126

\bibitem[\protect\citeauthoryear{Fahlman \& Fern\'andez}{Fahlman \& Fern\'andez}{2022}]{Fahlman:2022jkh}
Fahlman S.,  Fern\'andez R.,  2022, \mn@doi [Mon. Not. Roy. Astron. Soc.] {10.1093/mnras/stac948}, 513, 2689

\bibitem[\protect\citeauthoryear{Fern{\'a}ndez, Tchekhovskoy, Quataert, Foucart  \& Kasen}{Fern{\'a}ndez et~al.}{2019}]{Fernandez:2018kax}
Fern{\'a}ndez R.,  Tchekhovskoy A.,  Quataert E.,  Foucart F.,   Kasen D.,  2019, \mn@doi [Mon. Not. Roy. Astron. Soc.] {10.1093/mnras/sty2932}, 482, 3373

\bibitem[\protect\citeauthoryear{Fields, Prakash, Breschi, Radice, Bernuzzi  \& da Silva~Schneider}{Fields et~al.}{2023}]{Fields:2023}
Fields J.,  Prakash A.,  Breschi M.,  Radice D.,  Bernuzzi S.,   da Silva~Schneider A.,  2023, \mn@doi [Astrophys. J. Lett.] {10.3847/2041-8213/ace5b2}, 952, L36

\bibitem[\protect\citeauthoryear{Flanagan \& Hinderer}{Flanagan \& Hinderer}{2008}]{Flanagan:2007ix}
Flanagan E.~E.,  Hinderer T.,  2008, \mn@doi [Phys. Rev.] {10.1103/PhysRevD.77.021502}, D77, 021502

\bibitem[\protect\citeauthoryear{Foucart, M{\"o}sta, Ramirez, Wright, Darbha  \& Kasen}{Foucart et~al.}{2021}]{Foucart:2021}
Foucart F.,  M{\"o}sta P.,  Ramirez T.,  Wright A.~J.,  Darbha S.,   Kasen D.,  2021, \mn@doi [\prd] {10.1103/physrevd.104.123010}, 104, 1

\bibitem[\protect\citeauthoryear{{Freiburghaus}, {Rembges}, {Rauscher}, {Kolbe}, {Thielemann}, {Kratz}, {Pfeiffer}  \& {Cowan}}{{Freiburghaus} et~al.}{1999}]{Freiburghaus:1999a}
{Freiburghaus} C.,  {Rembges} J.~F.,  {Rauscher} T.,  {Kolbe} E.,  {Thielemann} F.~K.,  {Kratz} K.~L.,  {Pfeiffer} B.,   {Cowan} J.~J.,  1999, \mn@doi [Astrophys.\ J.] {10.1086/307072}, \href {https://ui.adsabs.harvard.edu/abs/1999ApJ...516..381F} {516, 381}

\bibitem[\protect\citeauthoryear{{Fujibayashi}, {Wanajo}, {Kiuchi}, {Kyutoku}, {Sekiguchi}  \& {Shibata}}{{Fujibayashi} et~al.}{2020}]{Fujibayashi:2020dvr}
{Fujibayashi} S.,  {Wanajo} S.,  {Kiuchi} K.,  {Kyutoku} K.,  {Sekiguchi} Y.,   {Shibata} M.,  2020, \mn@doi [Astrophys.\ J.] {10.3847/1538-4357/abafc2}, \href {https://ui.adsabs.harvard.edu/abs/2020ApJ...901..122F} {901, 122}

\bibitem[\protect\citeauthoryear{Fujibayashi, Kiuchi, Wanajo, Kyutoku, Sekiguchi  \& Shibata}{Fujibayashi et~al.}{2023}]{Fujibayashi:2022ftg}
Fujibayashi S.,  Kiuchi K.,  Wanajo S.,  Kyutoku K.,  Sekiguchi Y.,   Shibata M.,  2023, \mn@doi [Astrophys. J.] {10.3847/1538-4357/ac9ce0}, 942, 39

\bibitem[\protect\citeauthoryear{Grossman, Korobkin, Rosswog  \& Piran}{Grossman et~al.}{2014}]{Grossman:2013lqa}
Grossman D.,  Korobkin O.,  Rosswog S.,   Piran T.,  2014, \mn@doi [Mon. Not. Roy. Astron. Soc.] {10.1093/mnras/stt2503}, 439, 757

\bibitem[\protect\citeauthoryear{Haas et~al.,}{Haas et~al.}{2020}]{Haas:2020a}
Haas R.,  et~al., 2020, The {{Einstein Toolkit}}, \mn@doi{10.5281/ZENODO.4298887}

\bibitem[\protect\citeauthoryear{Hammond, Hawke  \& Andersson}{Hammond et~al.}{2021}]{Hammond:2021vtv}
Hammond P.,  Hawke I.,   Andersson N.,  2021, \mn@doi [Phys. Rev. D] {10.1103/PhysRevD.104.103006}, 104, 103006

\bibitem[\protect\citeauthoryear{Hebeler, Bogner, Furnstahl, Nogga  \& Schwenk}{Hebeler et~al.}{2011}]{Hebeler:2011}
Hebeler K.,  Bogner S.~K.,  Furnstahl R.~J.,  Nogga A.,   Schwenk A.,  2011, \mn@doi [Phys. Rev. C] {10.1103/PhysRevC.83.031301}, 83, 031301

\bibitem[\protect\citeauthoryear{Hinderer, Lackey, Lang  \& Read}{Hinderer et~al.}{2010}]{Hinderer:2009ca}
Hinderer T.,  Lackey B.~D.,  Lang R.~N.,   Read J.~S.,  2010, \mn@doi [Phys. Rev.] {10.1103/PhysRevD.81.123016}, D81, 123016

\bibitem[\protect\citeauthoryear{Horowitz et~al.,}{Horowitz et~al.}{2019}]{Horowitz:2019}
Horowitz C.~J.,  et~al., 2019, \mn@doi [Journal of Physics G: Nuclear and Particle Physics] {10.1088/1361-6471/ab0849}, 46, 083001

\bibitem[\protect\citeauthoryear{Hotokezaka \& Nakar}{Hotokezaka \& Nakar}{2020}]{Hotokezaka:2020}
Hotokezaka K.,  Nakar E.,  2020, \mn@doi [Astrophys. J.] {10.3847/1538-4357/ab6a98}, 891, 152

\bibitem[\protect\citeauthoryear{Hotokezaka, Kiuchi, Kyutoku, Okawa, Sekiguchi  et~al.}{Hotokezaka et~al.}{2013}]{Hotokezaka:2012ze}
Hotokezaka K.,  Kiuchi K.,  Kyutoku K.,  Okawa H.,  Sekiguchi Y.-i.,   et~al., 2013, \mn@doi [Phys. Rev.] {10.1103/PhysRevD.87.024001}, D87, 024001

\bibitem[\protect\citeauthoryear{Hotokezaka, Kyutoku, Sekiguchi  \& Shibata}{Hotokezaka et~al.}{2016}]{Hotokezaka:2016bzh}
Hotokezaka K.,  Kyutoku K.,  Sekiguchi Y.-i.,   Shibata M.,  2016, \mn@doi [Phys. Rev.] {10.1103/PhysRevD.93.064082}, D93, 064082

\bibitem[\protect\citeauthoryear{Jacobi, Guercilena, Huth, Ricigliano, Arcones  \& Schwenk}{Jacobi et~al.}{2023}]{Jacobi:2023}
Jacobi M.,  Guercilena F.~M.,  Huth S.,  Ricigliano G.,  Arcones A.,   Schwenk A.,  2023, \mn@doi [Mon. Not. Roy. Astron. Soc.] {10.1093/mnras/stad3738}, 527, 8812

\bibitem[\protect\citeauthoryear{Just, Bauswein, Pulpillo, Goriely  \& Janka}{Just et~al.}{2015}]{Just:2014fka}
Just O.,  Bauswein A.,  Pulpillo R.~A.,  Goriely S.,   Janka H.~T.,  2015, \mn@doi [Mon. Not. Roy. Astron. Soc.] {10.1093/mnras/stv009}, 448, 541

\bibitem[\protect\citeauthoryear{Just, Goriely, Janka, Nagataki  \& Bauswein}{Just et~al.}{2021}]{Just:2021cls}
Just O.,  Goriely S.,  Janka H.-T.,  Nagataki S.,   Bauswein A.,  2021, \mn@doi [Mon. Not. Roy. Astron. Soc.] {10.1093/mnras/stab2861}, 509, 1377

\bibitem[\protect\citeauthoryear{{Kasen} \& {Barnes}}{{Kasen} \& {Barnes}}{2019}]{Kasen:2018drm}
{Kasen} D.,  {Barnes} J.,  2019, \mn@doi [Astrophys.\ J.] {10.3847/1538-4357/ab06c2}, \href {https://ui.adsabs.harvard.edu/abs/2019ApJ...876..128K} {876, 128}

\bibitem[\protect\citeauthoryear{Kasen, Fern{\'a}ndez  \& Metzger}{Kasen et~al.}{2015}]{Kasen:2014toa}
Kasen D.,  Fern{\'a}ndez R.,   Metzger B.,  2015, \mn@doi [Mon. Not. Roy. Astron. Soc.] {10.1093/mnras/stv721}, 450, 1777

\bibitem[\protect\citeauthoryear{Kasen, Metzger, Barnes, Quataert  \& Ramirez-Ruiz}{Kasen et~al.}{2017}]{Kasen:2017}
Kasen D.,  Metzger B.,  Barnes J.,  Quataert E.,   Ramirez-Ruiz E.,  2017, \mn@doi [Nature] {10.1038/nature24453}, 551, 80

\bibitem[\protect\citeauthoryear{Kiuchi, Kyutoku, Shibata  \& Taniguchi}{Kiuchi et~al.}{2019}]{Kiuchi:2019lls}
Kiuchi K.,  Kyutoku K.,  Shibata M.,   Taniguchi K.,  2019, \mn@doi [Astrophys. J.] {10.3847/2041-8213/ab1e45}, 876, L31

\bibitem[\protect\citeauthoryear{{Lattimer} \& {Schramm}}{{Lattimer} \& {Schramm}}{1974}]{Lattimer:1974a}
{Lattimer} J.~M.,  {Schramm} D.~N.,  1974, \mn@doi [Astrophys.\ J.\ Lett.] {10.1086/181612}, \href {http://adsabs.harvard.edu/abs/1974ApJ...192L.145L} {192, L145}

\bibitem[\protect\citeauthoryear{Lattimer \& Swesty}{Lattimer \& Swesty}{1991}]{Lattimer:1991nc}
Lattimer J.~M.,  Swesty F.~D.,  1991, \mn@doi [Nucl. Phys.] {10.1016/0375-9474(91)90452-C}, A535, 331

\bibitem[\protect\citeauthoryear{Li \& Paczynski}{Li \& Paczynski}{1998}]{Li:1998bw}
Li L.-X.,  Paczynski B.,  1998, \mn@doi [Astrophys. J.] {10.1086/311680}, 507, L59

\bibitem[\protect\citeauthoryear{Lippuner \& Roberts}{Lippuner \& Roberts}{2015}]{Lippuner:2015gwa}
Lippuner J.,  Roberts L.~F.,  2015, \mn@doi [Astrophys. J.] {10.1088/0004-637X/815/2/82}, 815, 82

\bibitem[\protect\citeauthoryear{Lippuner \& Roberts}{Lippuner \& Roberts}{2017}]{Lippuner:2017tyn}
Lippuner J.,  Roberts L.~F.,  2017, \mn@doi [Astrophys. J. Suppl.] {10.3847/1538-4365/aa94cb}, 233, 18

\bibitem[\protect\citeauthoryear{Lodders, Palme  \& Gail}{Lodders et~al.}{2009}]{lodders2009AbundancesElementsSolar}
Lodders K.,  Palme H.,   Gail H.-P.,  2009, in Martienssen W.,  Tr{\"u}mper J.,  eds, , Vol.~4B, Solar {{System}}.
{Springer Berlin Heidelberg}, {Berlin, Heidelberg}, pp 712--770, \mn@doi{10.1007/978-3-540-88055-4_34}

\bibitem[\protect\citeauthoryear{Ludlam et~al.,}{Ludlam et~al.}{2022}]{Ludlam:2022}
Ludlam R.~M.,  et~al., 2022, \mn@doi [Astrophys. J.] {10.3847/1538-4357/ac5028}, 927, 112

\bibitem[\protect\citeauthoryear{Magistrelli, Bernuzzi, Perego  \& Radice}{Magistrelli et~al.}{2024}]{Magistrelli:2024}
Magistrelli F.,  Bernuzzi S.,  Perego A.,   Radice D.,  2024, preprint (ArXiv:2403.13883)

\bibitem[\protect\citeauthoryear{Margalit \& Metzger}{Margalit \& Metzger}{2017}]{Margalit:2017dij}
Margalit B.,  Metzger B.~D.,  2017, \mn@doi [Astrophys. J.] {10.3847/2041-8213/aa991c}, 850, L19

\bibitem[\protect\citeauthoryear{Martin, Perego, Arcones, Thielemann, Korobkin  \& Rosswog}{Martin et~al.}{2015}]{Martin:2015hxa}
Martin D.,  Perego A.,  Arcones A.,  Thielemann F.-K.,  Korobkin O.,   Rosswog S.,  2015, \mn@doi [Astrophys. J.] {10.1088/0004-637X/813/1/2}, 813, 2

\bibitem[\protect\citeauthoryear{Metzger}{Metzger}{2020}]{Metzger:2019zeh}
Metzger B.~D.,  2020, \mn@doi [Living Rev. Rel.] {10.1007/s41114-019-0024-0}, 23, 1

\bibitem[\protect\citeauthoryear{Metzger \& Fern\'{a}ndez}{Metzger \& Fern\'{a}ndez}{2014}]{Metzger:2014ila}
Metzger B.~D.,  Fern\'{a}ndez R.,  2014, \mn@doi [Mon. Not. Roy. Astron. Soc.] {10.1093/mnras/stu802}, 441, 3444

\bibitem[\protect\citeauthoryear{Metzger, Martinez-Pinedo, Darbha, Quataert, Arcones  et~al.}{Metzger et~al.}{2010}]{Metzger:2010sy}
Metzger B.,  Martinez-Pinedo G.,  Darbha S.,  Quataert E.,  Arcones A.,   et~al., 2010, \mn@doi [Mon. Not. Roy. Astron. Soc.] {10.1111/j.1365-2966.2010.16864.x}, 406, 2650

\bibitem[\protect\citeauthoryear{Miller et~al.}{Miller et~al.}{2019}]{Miller:2019cac}
Miller M.~C.,  et~al., 2019, \mn@doi [Astrophys. J.] {10.3847/2041-8213/ab50c5}, 887, L24

\bibitem[\protect\citeauthoryear{Miller et~al.}{Miller et~al.}{2021}]{Miller:2021qha}
Miller M.~C.,  et~al., 2021, \mn@doi [Astrophys. J. Lett.] {10.3847/2041-8213/ac089b}, 918, L28

\bibitem[\protect\citeauthoryear{M\"oller, Sierk, Ichikawa  \& Sagawa}{M\"oller et~al.}{2016}]{Moller:2015fba}
M\"oller P.,  Sierk A.~J.,  Ichikawa T.,   Sagawa H.,  2016, \mn@doi [Atom. Data Nucl. Data Tabl.] {10.1016/j.adt.2015.10.002}, 109-110, 1

\bibitem[\protect\citeauthoryear{Morozova, Piro, Renzo, Ott, Clausen, Couch, Ellis  \& Roberts}{Morozova et~al.}{2015}]{Morozova:2015bla}
Morozova V.,  Piro A.~L.,  Renzo M.,  Ott C.~D.,  Clausen D.,  Couch S.~M.,  Ellis J.,   Roberts L.~F.,  2015, \mn@doi [Astrophys. J.] {10.1088/0004-637X/814/1/63}, 814, 63

\bibitem[\protect\citeauthoryear{Most \& Raithel}{Most \& Raithel}{2021}]{Most:2021ktk}
Most E.~R.,  Raithel C.~A.,  2021, \mn@doi [Phys. Rev. D] {10.1103/PhysRevD.104.124012}, 104, 124012

\bibitem[\protect\citeauthoryear{Most, Papenfort, Dexheimer, Hanauske, Schramm, Stöcker  \& Rezzolla}{Most et~al.}{2019}]{Most:2018eaw}
Most E.~R.,  Papenfort L.~J.,  Dexheimer V.,  Hanauske M.,  Schramm S.,  Stöcker H.,   Rezzolla L.,  2019, \mn@doi [Phys. Rev. Lett.] {10.1103/PhysRevLett.122.061101}, 122, 061101

\bibitem[\protect\citeauthoryear{Nedora, Bernuzzi, Radice, Perego, Endrizzi  \& Ortiz}{Nedora et~al.}{2019}]{Nedora:2019jhl}
Nedora V.,  Bernuzzi S.,  Radice D.,  Perego A.,  Endrizzi A.,   Ortiz N.,  2019, \mn@doi [Astrophys. J.] {10.3847/2041-8213/ab5794}, 886, L30

\bibitem[\protect\citeauthoryear{Nedora, Radice, Bernuzzi, Perego, Daszuta, Endrizzi, Prakash  \& Schianchi}{Nedora et~al.}{2021a}]{Nedora:2021eojb}
Nedora V.,  Radice D.,  Bernuzzi S.,  Perego A.,  Daszuta B.,  Endrizzi A.,  Prakash A.,   Schianchi F.,  2021a, \mn@doi [Mon. Not. Roy. Astron. Soc.] {10.1093/mnras/stab2004}, 506, 5908

\bibitem[\protect\citeauthoryear{Nedora et~al.,}{Nedora et~al.}{2021b}]{Nedora:2020pak}
Nedora V.,  et~al., 2021b, \mn@doi [Astrophys. J.] {10.3847/1538-4357/abc9be}, 906, 98

\bibitem[\protect\citeauthoryear{Nedora et~al.,}{Nedora et~al.}{2022}]{Nedora:2020qtd}
Nedora V.,  et~al., 2022, \mn@doi [Class. Quant. Grav.] {10.1088/1361-6382/ac35a8}, 39, 015008

\bibitem[\protect\citeauthoryear{Neilsen, Liebling, Anderson, Lehner, O’Connor  et~al.}{Neilsen et~al.}{2014}]{Neilsen:2014hha}
Neilsen D.,  Liebling S.~L.,  Anderson M.,  Lehner L.,  O’Connor E.,   et~al., 2014, \mn@doi [Phys. Rev.] {10.1103/PhysRevD.89.104029}, D89, 104029

\bibitem[\protect\citeauthoryear{Panov, Freiburghaus  \& Thielemann}{Panov et~al.}{2001}]{Panov:2001}
Panov I.,  Freiburghaus C.,   Thielemann F.-K.,  2001, \mn@doi [Nucl. Phys. A] {https://doi.org/10.1016/S0375-9474(01)00797-7}, 688, 587

\bibitem[\protect\citeauthoryear{Perego, Rosswog, Cabez{\'o}n, Korobkin, K{\"a}ppeli, Arcones  \& Liebend{\"o}rfer}{Perego et~al.}{2014a}]{Perego:2014}
Perego A.,  Rosswog S.,  Cabez{\'o}n R.~M.,  Korobkin O.,  K{\"a}ppeli R.,  Arcones A.,   Liebend{\"o}rfer M.,  2014a, \mn@doi [Mon. Not. Roy. Astron. Soc.] {10.1093/mnras/stu1352}, 443, 3134

\bibitem[\protect\citeauthoryear{Perego, Gafton, Cabezón, Rosswog  \& Liebendörfer}{Perego et~al.}{2014b}]{Perego:2014qda}
Perego A.,  Gafton E.,  Cabezón R.,  Rosswog S.,   Liebendörfer M.,  2014b, \mn@doi [Astron. Astrophys.] {10.1051/0004-6361/201423755}, 568, A11

\bibitem[\protect\citeauthoryear{Perego, Radice  \& Bernuzzi}{Perego et~al.}{2017}]{Perego:2017wtu}
Perego A.,  Radice D.,   Bernuzzi S.,  2017, \mn@doi [Astrophys. J.] {10.3847/2041-8213/aa9ab9}, 850, L37

\bibitem[\protect\citeauthoryear{Perego, Bernuzzi  \& Radice}{Perego et~al.}{2019}]{Perego:2019adq}
Perego A.,  Bernuzzi S.,   Radice D.,  2019, \mn@doi [Eur. Phys. J.] {10.1140/epja/i2019-12810-7}, A55, 124

\bibitem[\protect\citeauthoryear{Perego, Logoteta, Radice, Bernuzzi, Kashyap, Das, Padamata  \& Prakash}{Perego et~al.}{2022a}]{Perego:2021mkd}
Perego A.,  Logoteta D.,  Radice D.,  Bernuzzi S.,  Kashyap R.,  Das A.,  Padamata S.,   Prakash A.,  2022a, \mn@doi [Phys. Rev. Lett.] {10.1103/PhysRevLett.129.032701}, 129, 032701

\bibitem[\protect\citeauthoryear{Perego et~al.}{Perego et~al.}{2022b}]{Perego:2020evn}
Perego A.,  et~al., 2022b, \mn@doi [Astrophys. J.] {10.3847/1538-4357/ac3751}, 925, 22

\bibitem[\protect\citeauthoryear{Pinto \& Eastman}{Pinto \& Eastman}{2000}]{Pinto:2000}
Pinto P.~A.,  Eastman R.~G.,  2000, \mn@doi [Astrophys. J.] {10.1086/308376}, 530, 744

\bibitem[\protect\citeauthoryear{Radice \& Rezzolla}{Radice \& Rezzolla}{2012}]{Radice:2012cu}
Radice D.,  Rezzolla L.,  2012, \mn@doi [Astron. Astrophys.] {10.1051/0004-6361/201219735}, 547, A26

\bibitem[\protect\citeauthoryear{Radice, Bernuzzi, Del~Pozzo, Roberts  \& Ott}{Radice et~al.}{2017}]{Radice:2016rys}
Radice D.,  Bernuzzi S.,  Del~Pozzo W.,  Roberts L.~F.,   Ott C.~D.,  2017, \mn@doi [Astrophys. J.] {10.3847/2041-8213/aa775f}, 842, L10

\bibitem[\protect\citeauthoryear{Radice, Perego, Bernuzzi  \& Zhang}{Radice et~al.}{2018a}]{Radice:2018xqa}
Radice D.,  Perego A.,  Bernuzzi S.,   Zhang B.,  2018a, \mn@doi [Mon. Not. Roy. Astron. Soc.] {10.1093/mnras/sty2531}, 481, 3670

\bibitem[\protect\citeauthoryear{Radice, Perego, Zappa  \& Bernuzzi}{Radice et~al.}{2018b}]{Radice:2017lry}
Radice D.,  Perego A.,  Zappa F.,   Bernuzzi S.,  2018b, \mn@doi [Astrophys. J.] {10.3847/2041-8213/aaa402}, 852, L29

\bibitem[\protect\citeauthoryear{Radice, Perego, Hotokezaka, Fromm, Bernuzzi  \& Roberts}{Radice et~al.}{2018c}]{Radice:2018pdn}
Radice D.,  Perego A.,  Hotokezaka K.,  Fromm S.~A.,  Bernuzzi S.,   Roberts L.~F.,  2018c, \mn@doi [Astrophys. J.] {10.3847/1538-4357/aaf054}, 869, 130

\bibitem[\protect\citeauthoryear{Radice, Perego, Hotokezaka, Fromm, Bernuzzi  \& Roberts}{Radice et~al.}{2018d}]{Radice:2018}
Radice D.,  Perego A.,  Hotokezaka K.,  Fromm S.~A.,  Bernuzzi S.,   Roberts L.~F.,  2018d, \mn@doi [Astrophys. J.] {10.3847/1538-4357/aaf054}, 869, 130

\bibitem[\protect\citeauthoryear{Raithel, Ozel  \& Psaltis}{Raithel et~al.}{2019}]{Raithel:2019gws}
Raithel C.~A.,  Ozel F.,   Psaltis D.,  2019, \mn@doi [Astrophys. J.] {10.3847/1538-4357/ab08ea}, 875, 12

\bibitem[\protect\citeauthoryear{Ravi \& Lasky}{Ravi \& Lasky}{2014}]{Ravi:2014}
Ravi V.,  Lasky P.~D.,  2014, \mn@doi [Mon. Not. Roy. Astron. Soc.] {10.1093/mnras/stu720}, 441, 2433

\bibitem[\protect\citeauthoryear{Reichert et~al.,}{Reichert et~al.}{2023}]{Reichert:2023}
Reichert M.,  et~al., 2023, \mn@doi [Astrophys. J. Suppl.] {10.3847/1538-4365/acf033}, 268, 66

\bibitem[\protect\citeauthoryear{Rezzolla, Most  \& Weih}{Rezzolla et~al.}{2018}]{Rezzolla:2017aly}
Rezzolla L.,  Most E.~R.,   Weih L.~R.,  2018, \mn@doi [Astrophys. J.] {10.3847/2041-8213/aaa401}, 852, L25

\bibitem[\protect\citeauthoryear{Ricigliano, Perego, Borhanian, Loffredo, Kawaguchi, Bernuzzi  \& Lippold}{Ricigliano et~al.}{2024}]{Ricigliano:2024}
Ricigliano G.,  Perego A.,  Borhanian S.,  Loffredo E.,  Kawaguchi K.,  Bernuzzi S.,   Lippold L.~C.,  2024, \mn@doi [Mon. Not. Roy. Astron. Soc.] {10.1093/mnras/stae572}, 529, 647

\bibitem[\protect\citeauthoryear{Riley et~al.}{Riley et~al.}{2019}]{Riley:2019yda}
Riley T.~E.,  et~al., 2019, \mn@doi [Astrophys. J.] {10.3847/2041-8213/ab481c}, 887, L21

\bibitem[\protect\citeauthoryear{Riley et~al.,}{Riley et~al.}{2021}]{Riley:2021}
Riley T.~E.,  et~al., 2021, \mn@doi [Astrophys. J. Lett.] {10.3847/2041-8213/ac0a81}, 918, L27

\bibitem[\protect\citeauthoryear{Rosswog, Sollerman, Feindt, Goobar, Korobkin, Wollaeger, Fremling  \& Kasliwal}{Rosswog et~al.}{2018}]{Rosswog:2017sdn}
Rosswog S.,  Sollerman J.,  Feindt U.,  Goobar A.,  Korobkin O.,  Wollaeger R.,  Fremling C.,   Kasliwal M.~M.,  2018, \mn@doi [Astron. Astrophys.] {10.1051/0004-6361/201732117}, 615, A132

\bibitem[\protect\citeauthoryear{Salmi et~al.}{Salmi et~al.}{2022}]{Salmi:2022cgy}
Salmi T.,  et~al., 2022, \mn@doi [Astrophys. J.] {10.3847/1538-4357/ac983d}, 941, 150

\bibitem[\protect\citeauthoryear{Schneider, Roberts  \& Ott}{Schneider et~al.}{2017}]{daSilvaSchneider:2017jpg}
Schneider A.~S.,  Roberts L.~F.,   Ott C.~D.,  2017, \mn@doi [Phys. Rev.] {10.1103/PhysRevC.96.065802}, C96, 065802

\bibitem[\protect\citeauthoryear{Sekiguchi, Kiuchi, Kyutoku, Shibata  \& Taniguchi}{Sekiguchi et~al.}{2016}]{Sekiguchi:2016bjd}
Sekiguchi Y.,  Kiuchi K.,  Kyutoku K.,  Shibata M.,   Taniguchi K.,  2016, \mn@doi [Phys. Rev.] {10.1103/PhysRevD.93.124046}, D93, 124046

\bibitem[\protect\citeauthoryear{Shen, Toki, Oyamatsu  \& Sumiyoshi}{Shen et~al.}{1998}]{Shen:1998gq}
Shen H.,  Toki H.,  Oyamatsu K.,   Sumiyoshi K.,  1998, \mn@doi [Nucl. Phys.] {10.1016/S0375-9474(98)00236-X}, A637, 435

\bibitem[\protect\citeauthoryear{Shibata, Fujibayashi, Hotokezaka, Kiuchi, Kyutoku, Sekiguchi  \& Tanaka}{Shibata et~al.}{2017}]{Shibata:2017xdx}
Shibata M.,  Fujibayashi S.,  Hotokezaka K.,  Kiuchi K.,  Kyutoku K.,  Sekiguchi Y.,   Tanaka M.,  2017, \mn@doi [Phys. Rev.] {10.1103/PhysRevD.96.123012}, D96, 123012

\bibitem[\protect\citeauthoryear{Shingles et~al.,}{Shingles et~al.}{2023}]{Shingles:2023kua}
Shingles L.~J.,  et~al., 2023, \mn@doi [Astrophys. J. Lett.] {10.3847/2041-8213/acf29a}, 954, L41

\bibitem[\protect\citeauthoryear{Sieverding, Waldrop, Harris, Hix, Lentz, Bruenn  \& Messer}{Sieverding et~al.}{2023}]{Sieverding:2023}
Sieverding A.,  Waldrop P.~G.,  Harris J.~A.,  Hix W.~R.,  Lentz E.~J.,  Bruenn S.~W.,   Messer O. E.~B.,  2023, \mn@doi [Astrophys. J.] {10.3847/1538-4357/acc8d1}, 950, 34

\bibitem[\protect\citeauthoryear{Smartt et~al.,}{Smartt et~al.}{2017}]{Smartt:2017}
Smartt S.~J.,  et~al., 2017, \mn@doi [Nature] {10.1038/nature24303}, 551, 75

\bibitem[\protect\citeauthoryear{Steiner, Hempel  \& Fischer}{Steiner et~al.}{2013}]{Steiner:2012rk}
Steiner A.~W.,  Hempel M.,   Fischer T.,  2013, \mn@doi [Astrophys. J.] {10.1088/0004-637X/774/1/17}, 774, 17

\bibitem[\protect\citeauthoryear{Tanaka \& Hotokezaka}{Tanaka \& Hotokezaka}{2013}]{Tanaka:2013ana}
Tanaka M.,  Hotokezaka K.,  2013, \mn@doi [Astrophys. J.] {10.1088/0004-637X/775/2/113}, 775, 113

\bibitem[\protect\citeauthoryear{{Tanaka}, {Kato}, {Gaigalas}  \& {Kawaguchi}}{{Tanaka} et~al.}{2020}]{Tanaka:2019iqp}
{Tanaka} M.,  {Kato} D.,  {Gaigalas} G.,   {Kawaguchi} K.,  2020, \mn@doi [Mon.\ Not.\ Roy.\ Astron.\ Soc.] {10.1093/mnras/staa1576}, \href {https://ui.adsabs.harvard.edu/abs/2020MNRAS.496.1369T} {496, 1369}

\bibitem[\protect\citeauthoryear{Tews, Margueron  \& Reddy}{Tews et~al.}{2018}]{Tews:2018chv}
Tews I.,  Margueron J.,   Reddy S.,  2018, \mn@doi [Phys. Rev.] {10.1103/PhysRevC.98.045804}, C98, 045804

\bibitem[\protect\citeauthoryear{Thielemann, Eichler, Panov  \& Wehmeyer}{Thielemann et~al.}{2017}]{Thielemann:2017acv}
Thielemann F.~K.,  Eichler M.,  Panov I.~V.,   Wehmeyer B.,  2017, \mn@doi [Ann. Rev. Nucl. Part. Sci.] {10.1146/annurev-nucl-101916-123246}, 67, 253

\bibitem[\protect\citeauthoryear{Watson et~al.}{Watson et~al.}{2019}]{Watson:2019xjv}
Watson D.,  et~al., 2019, \mn@doi [Nature] {10.1038/s41586-019-1676-3}, 574, 497

\bibitem[\protect\citeauthoryear{Winteler, Kaeppeli, Perego, Arcones, Vasset, Nishimura, Liebendoerfer  \& Thielemann}{Winteler et~al.}{2012}]{Winteler:2012hu}
Winteler C.,  Kaeppeli R.,  Perego A.,  Arcones A.,  Vasset N.,  Nishimura N.,  Liebendoerfer M.,   Thielemann F.-K.,  2012, \mn@doi [Astrophys. J. Lett.] {10.1088/2041-8205/750/1/L22}, 750, L22

\bibitem[\protect\citeauthoryear{Wollaeger et~al.,}{Wollaeger et~al.}{2018}]{Wollaeger:2017ahm}
Wollaeger R.~T.,  et~al., 2018, \mn@doi [Mon. Not. Roy. Astron. Soc.] {10.1093/mnras/sty1018}, 478, 3298

\bibitem[\protect\citeauthoryear{Wu, Ricigliano, Kashyap, Perego  \& Radice}{Wu et~al.}{2022}]{Wu:2021ibi}
Wu Z.,  Ricigliano G.,  Kashyap R.,  Perego A.,   Radice D.,  2022, \mn@doi [Mon. Not. Roy. Astron. Soc.] {10.1093/mnras/stac399}, 512, 328

\bibitem[\protect\citeauthoryear{Yasin, Sch{\"a}fer, Arcones  \& Schwenk}{Yasin et~al.}{2020}]{Yasin:2020a}
Yasin H.,  Sch{\"a}fer S.,  Arcones A.,   Schwenk A.,  2020, \mn@doi [Phys. Rev. Lett.] {10.1103/PhysRevLett.124.092701}, 124, 92701

\makeatother
\end{thebibliography}

%% ==============================================================
%%
%%                         APPENDIX
%%
%% ==============================================================
\appendix

\section{Tracers clustering and selection}
\label{app:tracer_selection}

\begin{figure}
    \includegraphics[width=\columnwidth]{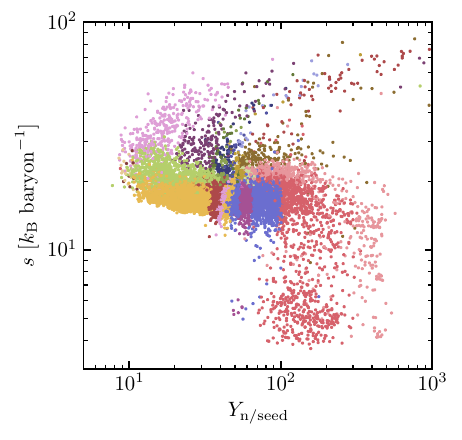}
    \caption{Example of tracers clustering based on final abundances.
    The tracers are obtained from the model \msS and are depicted as dots in the parameter space formed by the neutron-to-seed ratio and the specific entropy evaluated at $\sim5.8$ GK.
    Each color represents a different cluster, characterized by similar abundance patterns with less than one order of magnitude difference at the abundance peaks.}
    \label{fig:clusters}
\end{figure}

\begin{figure}
    \includegraphics[width=\columnwidth]{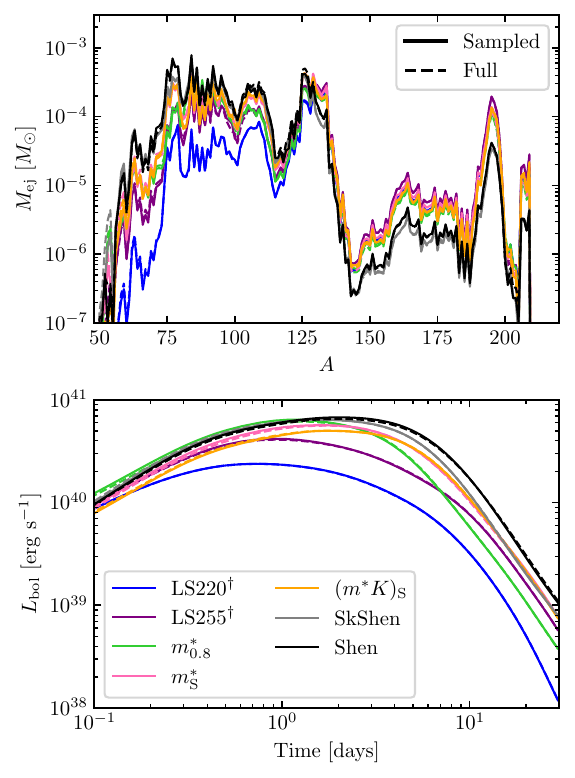}
    \caption{Final abundance pattern (top panel) and bolometric luminosity (bottom panel) for all the \ac{EOS} models, as computed using the entire set of extracted tracers and the subset obtained following the sampling procedure, which is fitted on the model \msS.}
    \label{fig:benchmark}
\end{figure}

In the context of explosive nucleosynthesis scenarios, the outflow hosting the nucleosynthesis process can be analysed through a number of tracers large enough in order to resolve all the relevant structural and compositional features.
Despite the relatively low computational cost of a single nuclear reaction network calculation, the latter is often repeated thousands to even millions of times for a complete outflow characterization.
However, among the many local thermodynamic initial conditions that are found within a single explosion event, some of them return similar if not nearly identical nucleosynthesis outcomes.
It is therefore possible to significantly reduce the amount of necessary information derived from hydrodynamical simulations, by identifying and grouping such conditions on the basis of the final yields \citep[see, e.g., ][for a similar analysis]{Bliss:2018nhk}.

In our approach, we first focus on a single simulation output (we choose the \msS model as representative) and we run the nuclear reaction network on the full collection of tracers.
The resulting final abundances are then analysed by employing a bottom-up hierarchical clustering procedure, in which we define a distance $d$ between two tracers $x$ and $y$ as:
\begin{equation}
    d_{x,y}\coloneqq\sum_{i:A_i>50}|{\rm log}(X_i^x)-{\rm log}(X_i^y)| \, ,
\end{equation}
where $X_i$ is the final mass fraction of the species $i$ and the sum runs over the species of interest for the r-process, with mass number $A_i>50$.
By tuning a minimum average distance between two separate clusters, we gather in the same group the abundance patterns that show analogous features, with less than one order of magnitude difference at the abundance peaks, and we obtain a few tens ($\sim35$) of separate clusters.

Subsequently, we choose a parameter space where these clusters live in connected and distinguishable regions.
In principle, the r-process nucleosynthesis is basically determined by the value of the neutron-to-seed ratio $Y_{\rm n/seed}=\frac{Y_{\rm n}}{Y_{\rm seed}}$ at the onset of the process, around $\sim3$ GK, with $Y_{\rm n}$ the abundance of free neutrons and $Y_{\rm seed}$ that of seed nuclei.
While at this temperature $Y_{\rm n/seed}$ can be exactly computed only using a reaction network, we approximately infer a representative $Y_{\rm n/seed}$ by evaluating it in \ac{NSE} at a higher temperature of $\sim5.8$ GK, commonly associated to \ac{NSE} freeze-out.
Despite, within this temperature shift, charged particle reactions can substantially alter the exact $Y_{\rm n/seed}$ value, we find that the error in this estimation does not qualitatively affect the clusters mapping.
We also find that clusters are well-reproduced when represented in a parameter space formed by the specific entropy $s$ at $5.8$ GK, together with $Y_{\rm n/seed}$, as shown in \reffig{fig:clusters} for our trial model.
In particular, tracers gather around a central region of interest with $Y_{\rm n/seed}\sim30-100$, where the highest variability of the nucleosynthesis outcomes is concentrated.

On the basis of this analysis, we construct a grid manually fitted to the cluster map, with a denser binning in the above inner region and a coarser binning in the outer regions, where abundances are less affected by the specific initial conditions.
For each of our models, we employ this grid in order to map every simulation tracer to the correspondent cluster.
We thus randomly select a variable number of representative tracers for each grid bin, and we use them to run our reduced nucleosynthesis calculations.
The remaining tracers are then associated to the nucleosynthesis evolution of the closer representative tracer in the parameter space.
We find that a 2D grid of 36 bins with 50 to 200 samples per bin is dense enough to reproduce the final results of each model with a good accuracy, as shown in \reffig{fig:benchmark}.
The obvious caveat to this approach is that, since its precision scales with the number of samples, local ejecta features can be under-resolved, especially in regions with an already low original number of tracers, e.g. in the low-density polar region.
Nevertheless, our findings suggest that global features of the ejecta, such as the total yields or the \ac{KN} emission, are the result of a broad average over all the different ejecta conditions, and as such they can be obtained already from a reduced amount of information.

\bsp    % typesetting comment
\label{lastpage}
\end{document}